\documentclass[draftcls,12pt,onecolumn]{IEEEtran}
\usepackage{bm}
\usepackage{amsmath, comment, color,soul,cite}
\usepackage{graphicx}
\usepackage{amsmath}
\usepackage{amssymb}
\usepackage{float}
\usepackage{acronym}
\usepackage{mymaths}
\usepackage{subfigure}
\usepackage{tikz}
\usepackage[normalem]{ulem}
\usetikzlibrary{dsp,chains}
\usetikzlibrary{arrows}
\usetikzlibrary{plotmarks}
\usepackage[tikz]{mypics}
\usepackage{psfrag}
\DeclareMathAlphabet{\mathpzc}{OT1}{pzc}{m}{it}

\acrodef{SNR}{signal-to-noise ratio}
\acrodef{i.i.d.}{independent identically distributed}
\acrodef{SVD}{singular value decomposition}
\acrodef{MAP}{maximum a posteriori}
\acrodef{MIMO}{multiple input multiple output}
\acrodef{OFDM}{orthogonal frequency division multiplexing}
\acrodef{CSI}{channel state information}
\acrodef{AWGN}{additive white Gaussian noise}
\acrodef{CDF}{cumulative distribution function}
\acrodef{KKT}{Karush-Kuhn-Tucker}
\acrodef{pdf}{probability density function}
\acrodef{BER}{bit error rate}
\acrodef{ML}{maximum likelihood}
\acrodef{SPB}{sphere packing bound}
\acrodef{CER}{codeword error rate}

\newcommand{\opt}{^\star}

\newcommand{\nota}[1]{{\slshape\color{blue}[#1]}}
\renewcommand{\nota}[1]{}

\newcommand{\myzeta}{\epsilon}
\newcommand{\myhl}[1]{#1}
\newcommand{\myhlF}[1]{#1}
\newcommand{\myhlFC}[1]{#1}
\newcommand{\myhlN}[1]{#1}

 \newtheorem{teo}{Theorem}

\renewcommand\P[1]{\mathbb{P}\left[#1\right]}

\newcommand{\baseplot}[3]{\draw[#1] plot file{Plots/#2/#3.table};}
\newcommand{\plotone}[2]{}
\newcommand{\plottwo}[2]{}

\newcommand{\legendone}[1]{} 

\usepackage[normalem]{ulem}
\usepackage{bm}
\usepackage{color}
\newcommand{\sot}[1]{}

\def\erreuno{R}

\usepackage[user]{zref}

\newcounter{revc}
\makeatletter \zref@newprop{revcontent}{} \zref@addprop{main}{revcontent}
\zref@newprop{revsec}{} \zref@addprop{main}{revsec}
\newcommand{\revi}[2]{%
\zref@setcurrent{revsec}{\thesection}%
\zref@setcurrent{revcontent}{#2}%
\refstepcounter{revc}%
\label{#1}
\zlabel{#1}%
#2 }
\newcommand{\revr}[2]{%
\zref@setcurrent{revsec}{\thesection}%
\zref@setcurrent{revcontent}{#2}%
\refstepcounter{revc}%
\zlabel{#1}%
\label{#1} \sot{#2}} \makeatother

\begin{document}
\sloppy
\title{Secrecy Transmission on Parallel Channels: Theoretical Limits and Performance\\  of Practical Codes}
%\author{M. Baldi, M. Bianchi, F. Chiaraluce, N. Laurenti, F. Renna, and S. Tomasin}

\author{Marco Baldi,~\IEEEmembership{Senior Member,~IEEE,} Franco Chiaraluce,~\IEEEmembership{Member,~IEEE,} Nicola Laurenti, \\ Stefano Tomasin,~\IEEEmembership{Senior Member,~IEEE,} and Francesco Renna,~\IEEEmembership{Member,~IEEE}%
\thanks{Manuscript received ***; revised ***.}%
\thanks{M. Baldi and F. Chiaraluce are with the Dipartimento di Ingegneria dell'Informazione, Universit\`a Politecnica delle Marche, 60131 Ancona, Italy.  N. Laurenti and S. Tomasin are with the Department of Information Engineering, University of Padua, 35131 Padua, Italy. F. Renna is with the Instituto de Telecomunica\c{c}\~{o}es e Departamento de Ci\^{e}ncia de Computadores, Faculdade de Ci\^{e}ncias da Universidade do Porto, 4169-007, Porto, Portugal}
\thanks{This work was supported in part by the Italian Ministry of Education and Research (MIUR) project ESCAPADE (Grant RBFR105NLC) under the ``FIRB-Futuro in Ricerca 2010'' funding program.}
}

\maketitle
\acresetall

\markboth{Transactions on Information Forensics and Security,~Vol.~1, No.~1,~January~2025}{Baldi \MakeLowercase{\text it{et al.}}: Secrecy Transmission on Parallel Channels}

\IEEEpubid{0000--0000/00\$00.00~\copyright~2014 IEEE}

%\vspace{-1em}
\begin{abstract}
We consider a system where an agent (Alice) aims at transmitting a
message to a second agent (Bob) over a set of parallel channels, while 
keeping it secret from a third agent (Eve) by using physical layer security techniques. 
We assume that Alice
perfectly knows the set of channels with respect to Bob, but she has only a
statistical knowledge of the channels with respect to Eve. We derive
bounds on the  achievable outage secrecy rates, by considering coding either
within each channel or across all parallel channels.
Transmit power is adapted to the channel conditions, with a constraint on
the average power over the whole transmission. We also focus on the
maximum cumulative outage secrecy rate that can be achieved. Moreover, in order to assess
the performance in a real life scenario, we consider the use of
practical error correcting codes.
We extend the definitions of security gap and equivocation rate,
previously applied to the single additive white Gaussian noise channel, 
to Rayleigh distributed parallel channels, on the basis of the error rate targets
and the outage probability.
Bounds on these metrics are also derived, taking into account
the statistics of the parallel channels.
Numerical results are provided, that confirm the
feasibility of the considered physical layer security techniques.
\end{abstract}

%\newpage
\begin{IEEEkeywords} Coding, outage probability, parallel channels, physical layer security. \end{IEEEkeywords}

%\clearpage
\acresetall
\section{Introduction}
\label{sec:one}

\IEEEPARstart{P}{erformance} of physical layer security schemes can be assessed either by evaluating their achievable secrecy rates
%\footnote{For physical layer security, the secrecy rate is defined as the rate at which information can be transmitted secretly from the source to the legal destination; its maximum achievable value is the secrecy capacity.} 
-- which assume, among other things, ideal coding (e.g., Gaussian codewords with infinite length) -- or focusing on practical codes and considering the error probabilities for both the legitimate receiver and the eavesdropper.

Within the former approach, the ergodic secrecy capacity for a fast fading scenario is derived in {\cite{Gopala08}} by maximizing the ergodic secrecy rate over all power allocations that meet an average transmit power constraint.
A compound parallel Gaussian wiretap channel, in which the main channel gains are known to all parties, while the eavesdropper gains can take any value within a given finite set, is considered in {\cite{LiuISIT08}}, where a max-min coding strategy is proved to achieve secrecy capacity. In the block fading scenario of {\cite{TangCISS07}}, only statistics of both the legitimate receiver and the eavesdropper channels are assumed to be known at the transmitter.
Then, a secrecy throughput is evaluated, that is achieved either with repetition coding or with a single wiretap channel code over a finite number of fading blocks. On the other hand, it is necessary to take into account the probability that,
for a certain fraction of time, the transmission becomes either unreliable (reliability outage) or insecure (secrecy outage). The statistical distribution of the secrecy capacity and the low-rate limit on the secrecy outage probability are derived in {\cite{Renna12,Renna10}} for a set of independently faded parallel wiretap channels, thus modeling \ac{OFDM} transmissions.
In {\cite{LinPH12}}, perfect \mbox{\ac{CSI}} for the main channel and statistical \mbox{\ac{CSI}} for the eavesdropper channel are assumed: for a fast Rayleigh fading wiretap channel with a multi-antenna transmitter and a single antenna device for both the intended receiver and the eavesdropper (MISOSE channel), the ergodic secrecy rate is optimized through an artificial noise injection scheme.  Similarly, {\cite{LinSC12}} fully characterizes the ergodic secrecy capacity of the MISOSE channel under statistical \mbox{\ac{CSI}} for both the main and the eavesdropper channels.

Very few examples of practical codes over (different kinds of) wiretap channels have been studied in previous literature. Most of these papers aim at finding codes able to achieve the secrecy capacity. This problem has been solved for the binary erasure channel (BEC), where low-density parity-check (LDPC) codes have been considered {\cite{Suresh2010}}, and for the binary symmetric channel (BSC), where polar codes have been proposed {\cite{Mahdavifar2011}}.
More recently, polar codes have also been included in a key agreement protocol over block fading channels {\cite{Koyluoglu2012}}. 
Polar coding, however, has been shown to be optimal over discrete memoryless channels, while our focus is on continuous-output channel models, which are best suited to model wireless transmissions.

\IEEEpubidadjcol 
To the best of the authors knowledge, at this time no code is available that ensures \myhlN{information theoretic secrecy, even asymptotically (e.g., in one of the criteria listed in \cite{BlochLaneman08})} over continuous output channels; therefore, other secrecy metrics must be considered. A first step toward practical scenarios is provided by the equivocation rate {\cite{WongWong2011}}, that still considers information leakage as a security metric, while taking into account the rate that can be reliably decoded by Bob with practical codes. However, it is still assumed that Eve may get an information rate equal to her channel capacity. When this assumption is removed and the error rate that can be achieved even by Eve is taken into account, an interesting metric is the {\em security gap} {\cite{Klinc2009a}}, that compares the \mbox{\acp{SNR}} on the main and the eavesdropper channels required to achieve both a sufficient level of secrecy and reliable decoding by the authorized receiver. In other words, the security gap is the required legitimate receiver power margin for having both a sufficiently high probability that he correctly receives the transmitted message and a sufficiently high probability that the message is not gathered by an eavesdropper. The security gap metric has been applied in {\cite{Klinc2011}} to punctured LDPC codes and in {\cite{Baldi2010, Baldi2011, Baldi2012}} to non-systematic codes, including LDPC codes and classical Bose-Chaudhuri-Hocquenghem (BCH) codes. Practical codes for physical layer security have also been applied over the packet erasure channel {\cite{Harrison2011}}, where some properties of stopping sets are exploited to achieve secrecy with punctured non-systematic LDPC codes. However, to the best of our knowledge, the evaluation of secrecy capabilities, with the security gap metric, for practical codes over parallel channels has never been faced in previous literature.

In this paper, we consider a parallel channels scenario where a transmitter, Alice, and a legitimate receiver, Bob, have perfect \ac{CSI} for their link, while Alice only has a statistical description of the link between herself and the eavesdropper, Eve. Both the Alice-Bob and the Alice-Eve links are assumed to be \ac{i.i.d.} Rayleigh parallel channels, modeling for example an \ac{OFDM} transmission over independently faded subcarriers. As the channel gains are represented by continuous random variables, the compound parallel Gaussian wiretap channel model {\cite{LiuISIT08}} does not apply. Moreover, transmission is performed over a finite set of parallel channels, thus preventing the leverage of ergodicity for the fading wiretap channel {\cite{Gopala08}}. Therefore, the transmission scenario implies a nonzero secrecy outage probability {\cite{Zhou11}}, and we aim at maximizing the secrecy rates while satisfying a constraint on the secrecy outage probability. 

Two approaches are considered for the transmission of a message by Alice: in one case, the message is first split into sub-messages, each separately encoded and transmitted on a different channel; in the other case, the message is encoded into a single codeword which is split into sub-words, each transmitted on a different channel. The first case is denoted as {\em coding per sub-message} (CPS), while the second one is denoted as {\em coding across sub-messages} (CAS).
This distinction is similar to the one between \emph{variable} and \emph{constant rate} transmission in {\cite{Gopala08}}. However, in \cite{Gopala08} codewords are assumed to span all the possible fading states, thus reducing it to an ergodic scenario.  Here instead we consider a finite set of parallel channels, thus taking into account the possibility of secrecy outage. The performance of the proposed scheme is assessed both by information theoretical arguments and by evaluation of the error rates with existing practical codes.

The main contributions of the paper are:
\begin{itemize}
\item the derivation of achievable secrecy rates for transmissions over independent Rayleigh distributed parallel channels subject to a secrecy outage probability constraint;
\item the joint optimization of power and rate allocation among sub-messages for secrecy rate maximization subject to a constraint on the maximum secrecy outage probability, where
the compound parallel Gaussian wiretap channel model {\cite{LiuISIT08}} cannot be applied;
\item the derivation of closed-form expressions of the outage secrecy rates for both CPS and CAS scenarios, otherwise previously available only by Monte Carlo methods {\cite{TangCISS07}};
\item the non-trivial performance comparison between CPS and CAS, since their outage secrecy rates are not immediately comparable;
\item the derivation of bounds on the error rates for Bob and Eve with practical codes on Rayleigh distributed parallel channels;
\item the extension of the security gap and the equivocation rate metrics from a single \ac{AWGN} channel to Rayleigh distributed parallel channels.
\end{itemize}
 
The paper is organized as follows. In Section \ref{sec:two} we introduce the system model, and in Section \ref{sec:three} we derive theoretical bounds on the achievable rates, for both CPS and CAS. In Section \ref{sec:four} we use the error rate as a different metric to assess the physical layer security on parallel channels, when practical codes are applied. Section \ref{sec:five}  provides several numerical examples, and Section \ref{sec:six} concludes the paper.

\section{System Model}
\label{sec:two}

%We consider a scenario with three agents: Alice, Bob, and Eve. Alice aims at transmitting a secret message to Bob, while Eve attempts to intercept it. 

\myhl{
Let us consider a scenario with $K$ parallel wiretap channels with independent Rayleigh distributed fading and  \ac{AWGN}. We denote by $h_k$ the complex (baseband equivalent) channel coefficient between Alice and Bob upon transmission over the channel $k = 1, 2, \ldots, K$, and $g_k$ the corresponding channel coefficient between Alice and Eve. Both coefficients are assumed to be constant for the duration of a transmission. The power gains  $H_k = |h_k|^2$, and $G_k = |g_k|^2$ are independent exponentially distributed random variables with means $\alpha_{\rm B}$ and $\alpha_{\rm E}$, respectively. The thermal noise variance of all channels is normalized to one.
Let us define the vectors \myhlF{$\bm{H}=[H_1,\ldots,H_K]$ and} $\bm{P} = [P_1, \ldots, P_K]$, where $P_k$ is the power transmitted by Alice over the $k$-th channel.

In the following, we will refer to the notion of \emph{secrecy rate}, that is a transmission rate for which, in the asymptotic regime of infinite codeword length, it is possible to guarantee that the error probability at Bob's receiver approaches zero (\emph{reliability condition}) and that mutual information between the transmitted message and the signal received by Eve is arbitrarily small (\myhlN{\emph{strong secrecy condition}, as formally defined in, e.g., \cite[Sec. 3.3]{Bloch}}). Secrecy and reliability can be ensured if Alice has \ac{CSI} on channels to both Bob and Eve. However, as better explained in the following, Alice is assumed to know only the statistical description of the channel coefficients $g_k$. In this case, using a code with a given secrecy rate may lead to some information leakage to Eve (depending on channel conditions), i.e., to a \emph{secrecy outage}~{\cite{Zhou11}} event.
}

\begin{figure}
\begin{centering}
\begin{footnotesize}
\begin{tikzpicture}
\def\radius{1mm} 
\tikzstyle{branch}=[fill,shape=circle,minimum size=3pt,inner sep=0pt]

	% Place nodes using a matrix
	\matrix (m1) [row sep=1mm, column sep=1.6mm]
	{
		%--------------------------------------------------------------------
		\node[dspnodeopen,dsp/label=above] (m00) {Alice};    &
		\node[coordinate] (m0ad) {};    &
		\node[dspfilter]  (m01) {\,S/P\, }; & &
		\node[dspfilter]  (m02) {\,ENC$_1$\,}; &
		\node[dspmixer]   (m03) {};    &		
		\node[dspnodefull](m04) {};          &
		\node[coordinate](m05) {};          &				
		\node[dspmixer]   (m06) {};    &
		\node[dspadder]   (m07) {};    &
		\node[dspfilter]  (m08) {\,DEC$_1^{\rm B}$\,}; & &
		\node[dspfilter]  (m09) {\,P/S\, }; &		&
		\node[dspnodeopen,dsp/label=above] (m010) {Bob};  
\\
& & & & & \node (E) {$\sqrt{P_1}$};
 & & & \node (A) {$h_1$};
 & \node (C) {noise}; & &
\\[-3mm]
& & & & \node (pee) {$\vdots$}; & \node (pee) {$\vdots$}; 
 & & & \node (pee) {$\vdots$}; 
 & \node (pee) {$\vdots$}; & &
\\
		%--------------------------------------------------------------------
&		\node[coordinate] (m10) {};    &
		\node[coordinate] (m11) {};    &	&	
		\node[dspfilter]  (m12) {\,ENC$_K$\,}; &
		\node[dspmixer]   (m13) {};    &		
		\node[coordinate] (m14) {};          &
		\node[coordinate] (m15) {};          &		
		\node[dspmixer, dsp/label=below]   (m16) {};    &
		\node[dspadder]   (m17) {};    &
		\node[dspfilter]  (m18) {\,DEC$_K^{\rm B}$\,}; &	&		
		\node[coordinate] (m19) {}; &   
\\
& & & & & \node (E2) {$\sqrt{P_K}$};
 & & & \node (B) {$h_K$};
 & \node (D) {noise}; & &&
\\[2mm]
		%--------------------------------------------------------------------
& 		\node[coordinate] (p00) {};    &
		&
		&
		&	&	
		\node[coordinate](p04) {};          &
		\node[coordinate](p05) {};          &				
		\node[dspmixer]   (p06) {};    &
		\node[dspadder]   (p07) {};    &
		\node[dspfilter]  (p08) {\,DEC$_1^{\rm E}$\,}; &&		
		\node[dspfilter]  (p09) {\,P/S\, }; &		&
		\node[dspnodeopen,dsp/label=above] (p010) {Eve};   
\\
& & & & &
& & & \node (pA) {$g_1$};
& \node (pC) {noise}; & & 
\\[-3mm]
& & & 
& & & \node (pee) {$\vdots$}; 
& \node (pee1) {$\vdots$}; & & 
\\
%--------------------------------------------------------------------
		& &
		&		
		&
		&		
		&&
		\node[coordinate] (p15) {};          &		
		\node[dspmixer, dsp/label=below]   (p16) {};    &
		\node[dspadder]   (p17) {};    &
		\node[dspfilter]  (p18) {\,DEC$_K^{\rm E}$\,}; &	&		
		\node[coordinate] (p19) {}; &   
\\
& & & & &
& & & \node (pB) {$g_K$};
& \node (pD) {noise}; & & &
\\
	};

	% Draw connections

	\foreach \i [evaluate = \i as \j using int(\i),
	             evaluate = \i as \k using int(\i+1),] in {0,2,6,7}
	{
		\begin{scope}[start chain]
			\chainin (m0\i);
			\chainin (m0\k) [join=by dspconn];
		\end{scope}
	}
		\begin{scope}[start chain]
			\chainin (m09);
			\chainin (m010) [join=by dspconn];
		\end{scope}
		\begin{scope}[start chain]
			\chainin (m03);
			\chainin (m06) [join=by dspconn];
		\end{scope}

   \draw[dspconn] (m01.20) node[above] {}{} |- (m02.160);
   \draw[dspline] (m01.-20)+(.1, 0) node {}{} -- (m01.-20)+(.4, 0);           
   \draw[dspconn] (m01.-20)+(.1, 0) node[above] {}{} |- (m12.west);

\path [draw, dspconn] (m08.20) -- node [above] {} 
        (m09.160) ;
\path [draw, dspline] (m18.0)+(.1,0) -- 	(m18.0)+(.4,0) ;
   \draw[dspconn] (m18.east)+(.1,0) node {}{} |- (m09.200);

	\foreach \i [evaluate = \i as \j using int(\i),
	             evaluate = \i as \k using int(\i+1),] in {2,6,7}
	{

		\begin{scope}[start chain]
			\chainin (m1\i);
			\chainin (m1\k) [join=by dspconn];
		\end{scope}

	}
		\begin{scope}[start chain]
			\chainin (m13);
			\chainin (m16) [join=by dspconn];
		\end{scope}

	\foreach \i [evaluate = \i as \j using int(\i),
	             evaluate = \i as \k using int(\i+1),] in {6,7}
	{
		\begin{scope}[start chain]
			\chainin (p0\i);
			\chainin (p0\k) [join=by dspconn];
		\end{scope}
	}
		\begin{scope}[start chain]
			\chainin (p09);
			\chainin (p010) [join=by dspconn];
		\end{scope}
%		\begin{scope}[start chain]
%			\chainin (p04);
%			\chainin (p06) [join=by dspconn];
%		\end{scope}

\path [draw, dspconn] (p08.20) -- 	(p09.160) ;	
\path [draw, dspline] (p18.0)+(.1,0) -- 	(p18.0)+(.4,0) ;	
\draw[dspconn] (p18.east)+(.1,0) node {}{} |- (p09.200);      
                      
	\foreach \i [evaluate = \i as \j using int(\i),
	             evaluate = \i as \k using int(\i+1),] in {6,7}
	{

		\begin{scope}[start chain]
			\chainin (p1\i);
			\chainin (p1\k) [join=by dspconn];
		\end{scope}

	}
\draw[dspconn] (E) -- (m03);
\draw[dspconn] (E2) -- (m13);
\draw[dspconn] (A) -- (m06);
\draw[dspconn] (B) -- (m16);
\draw[dspconn] (C) -- (m07);
\draw[dspconn] (D) -- (m17);

\draw[dspconn] (pA) -- (p06);
\draw[dspconn] (pB) -- (p16);
\draw[dspconn] (pC) -- (p07);
\draw[dspconn] (pD) -- (p17);

    \tikzstyle{s}=[shift={(0mm,\radius)}]
   \draw[dspconn] (m04) node[branch] {}{ % draw branch junction
                [shift only] -- ([s]m14 -| m04) arc(90:-90:\radius)
        } |- (p06);
   \draw[dspconn] (m15) node[branch] {}{ % draw branch junction
                [shift only] -- ([s]p05 -| m15) arc(90:-90:\radius)
        } |- (p16);
        
\end{tikzpicture}\end{footnotesize}
\caption{System model for CPS. S/P: serial to parallel; P/S: parallel to serial.}
\label{figmodCPS}
\par\end{centering}
\end{figure}
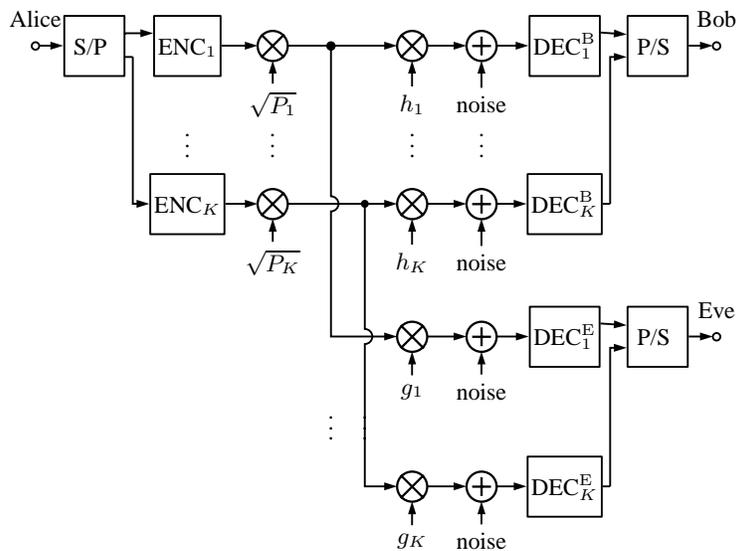

As stated in Section \ref{sec:one}, two coding approaches are considered:

\paragraph*{Coding per sub-message (CPS)} In this case, Alice first splits the message into $K$ sub-messages; each of them is then encoded into a different codeword and transmitted over a different channel, as shown in Fig. \ref{figmodCPS}. The secrecy rate for sub-message $k =1, 2, \ldots, K$ is $R_k$.
In other terms, each sub-message is encoded independently using a wiretap channel code with secrecy rate $R_k$, where $\sum_{k=1}^K R_k$ is the \myhlN{total message rate}.

\begin{figure}
\centering\begin{footnotesize}
\begin{tikzpicture}
\def\radius{1mm} 
\tikzstyle{branch}=[fill,shape=circle,minimum size=3pt,inner sep=0pt]

	% Place nodes using a matrix
	\matrix (m1) [row sep=1mm, column sep=1.6mm]
	{
		%--------------------------------------------------------------------
		\node[dspnodeopen,dsp/label=above] (m00) {Alice}; &
		\node[coordinate] (m0ad) {};    &		
		\node[dspfilter]  (m01) {\,ENC\, }; &
		\node[dspfilter]  (m02) {\,S/P\, }; &
		\node[dspmixer]   (m03) {}; &		
		\node[dspnodefull](m04) {}; &
		\node[coordinate] (m05) {}; & 
		\node[dspmixer]   (m06) {};    &
		\node[dspadder]   (m07) {};    &
		\node[dspfilter]  (m08) {\,P/S\,}; &		
		\node[dspfilter]  (m09) {\,DEC$^{\rm B}$\, }; &	&	
		\node[dspnodeopen,dsp/label=above] (m010) {Bob};    
\\
& &  & & \node (E) {$\sqrt{P_1}$};
& & & \node (A) {$h_1$};
& \node (C) {noise\;}; & &
\\[-3mm]
& & & \node (pee2) {$\vdots$}; 
& & & \node (pee3) {$\vdots$}; 
& \node (pee3) {$\vdots$}; & & 
\\	%--------------------------------------------------------------------
&		\node[coordinate] (m10) {}; &
		\node[coordinate] (m11) {}; &		
		\node[coordinate] (m12) {}; &
		\node[dspmixer]   (m13) {}; &		
		\node[coordinate] (m14) {}; &
		\node[coordinate] (m15) {}; &		
		\node[dspmixer]   (m16) {}; &
		\node[dspadder]   (m17) {}; &
		\node[coordinate]  (m18){}; &			
		\node[coordinate] (m19) {};    
\\[2mm]
& & & & \node (E2) {$\sqrt{P_K}$};
& & & \node (B) {$h_K$};
& \node (D) {noise}; & &
\\[2mm]
%--------------------------------------------------------------------
&		\node[coordinate] (p00) {};    &
		&
		&
		&		
		\node[coordinate](p04) {};          &
		\node[coordinate](p05) {};          &				
		\node[dspmixer]   (p06) {};    &
		\node[dspadder]   (p07) {};    &
		\node[dspfilter]  (p08) {\,P/S\,}; &		
		\node[dspfilter]  (p09) {\,DEC$^{\rm E}$\, }; &	&	
		\node[dspnodeopen,dsp/label=above] (p010) {Eve};   
		\\
& & & & 
& & & \node (pA) {$g_1$};
& \node (pC) {noise\;}; & &
\\[-3mm]
& & & & 
& & & \node (pee) {$\vdots$}; 
& \node (pee1) {$\vdots$}; & & 
\\
%--------------------------------------------------------------------
&		&
		&		
		&
		&		
		&
		\node[coordinate] (p15) {};          &		
		\node[dspmixer, dsp/label=below]   (p16) {};    &
		\node[dspadder]   (p17) {};    &
		\node[coordinate]  (p18) {}; &			
		\node[coordinate] (p19) {};  
\\[2mm]
& & & & 
& & & \node (pB) {$g_K$};
& \node (pD) {noise}; & &
\\
	};

	% Draw connections

	\foreach \i [evaluate = \i as \j using int(\i),
	             evaluate = \i as \k using int(\i+1),] in {0,1,2,6,7}
	{
		\begin{scope}[start chain]
			\chainin (m0\i);
			\chainin (m0\k) [join=by dspconn];
		\end{scope}
	}
		\begin{scope}[start chain]
			\chainin (m09);
			\chainin (m010) [join=by dspconn];
		\end{scope}
		\begin{scope}[start chain]
			\chainin (m03);
			\chainin (m06) [join=by dspconn];
		\end{scope}

   \draw[dspline] (m02.-20)+(.1, 0) node {}{} -- (m02.-20)+(.4, 0);           
   \draw[dspconn] (m02.-20)+(.1, 0) node[above] {}{} |- (m13.west);

\path [draw, dspconn] (m08) -- node [above] {} 
        (m09) ;
\path [draw, dspline] (m17.0)+(.1,0) -- 	(m17.0)+(.4,0) ;
   \draw[dspconn] (m17.east)+(.1,0) node {}{} |- (m08.200);

	\foreach \i [evaluate = \i as \j using int(\i),
	             evaluate = \i as \k using int(\i+1),] in {6}
	{

		\begin{scope}[start chain]
			\chainin (m1\i);
			\chainin (m1\k) [join=by dspconn];
		\end{scope}

	}
		\begin{scope}[start chain]
			\chainin (m13);
			\chainin (m16) [join=by dspconn];
		\end{scope}

	\foreach \i [evaluate = \i as \j using int(\i),
	             evaluate = \i as \k using int(\i+1),] in {6,7}
	{
		\begin{scope}[start chain]
			\chainin (p0\i);
			\chainin (p0\k) [join=by dspconn];
		\end{scope}
	}
		\begin{scope}[start chain]
			\chainin (p09);
			\chainin (p010) [join=by dspconn];
		\end{scope}
%		\begin{scope}[start chain]
%			\chainin (p04);
%			\chainin (p06) [join=by dspconn];
%		\end{scope}

\path [draw, dspconn] (p08) -- 	(p09) ;	
\path [draw, dspline] (p17.0)+(.1,0) -- 	(p17.0)+(.4,0) ;	
\draw[dspconn] (p17.east)+(.1,0) node {}{} |- (p08.200);      
                      
	\foreach \i [evaluate = \i as \j using int(\i),
	             evaluate = \i as \k using int(\i+1),] in {6}
	{

		\begin{scope}[start chain]
			\chainin (p1\i);
			\chainin (p1\k) [join=by dspconn];
		\end{scope}

	}
\draw[dspconn] (E) -- (m03);
\draw[dspconn] (E2) -- (m13);
\draw[dspconn] (A) -- (m06);
\draw[dspconn] (B) -- (m16);
\draw[dspconn] (C) -- (m07);
\draw[dspconn] (D) -- (m17);

\draw[dspconn] (pA) -- (p06);
\draw[dspconn] (pB) -- (p16);
\draw[dspconn] (pC) -- (p07);
\draw[dspconn] (pD) -- (p17);

    \tikzstyle{s}=[shift={(0mm,\radius)}]
   \draw[dspconn] (m04) node[branch] {}{ % draw branch junction
                [shift only] -- ([s]m14 -| m04) arc(90:-90:\radius)
        } |- (p06);
   \draw[dspconn] (m15) node[branch] {}{ % draw branch junction
                [shift only] -- ([s]p05 -| m15) arc(90:-90:\radius)
        } |- (p16);
        
\end{tikzpicture}\end{footnotesize}

\caption{System model for CAS.}
\label{figmodCAS}
\end{figure}
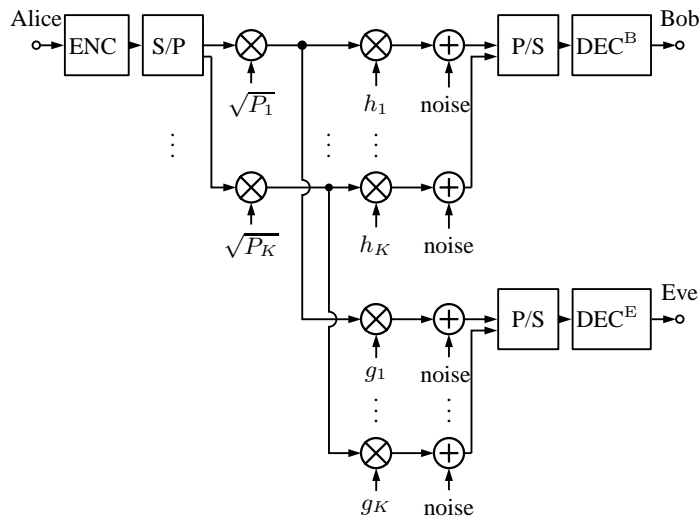

\paragraph*{Coding across sub-messages (CAS)} In this case, the message is first encoded into a single codeword and then transmitted over the $K$ parallel channels, as shown in Fig. \ref{figmodCAS}. The secrecy rate of the message is denoted as $\erreuno$.

\myhlF{
%For example, in a conventional point to point transmission without secrecy constraints the two schemes achieve the same rates} (\textbf{WE NEED A REFERENCE HERE}). \marker{Questa cosa non mi convince, parliamone!} \marker{Secondo me confondiamo `full CSI vs statistical CSI', e `secret vs unconstrained', e `ergodic vs finite set of realizations'.} \myhlFC{On the other hand}, when security is taken into account the two schemes \myhlFC{may be} not anymore equivalent. %\myhlFC{In particular}, in {\cite{Gopala08}} it is shown that CPS (in that paper denoted as variable rate coding scheme) achieves a higher secrecy rate than CAS (there denoted as constant rate coding scheme) in an ergodic scenario of infinitely long transmissions spanning the entire channel statistics. 
In this paper we investigate the two schemes when a transmission spans a finite set of channel realizations and Alice does not have \ac{CSI} of the channel to Eve. Then, secrecy conditions may be not satisfied and the outage probability is considered as a performance metric together with the secrecy rate. We will see that also in this scenario the performance of the two schemes differ. 

We first observe that CPS is a special case of CAS, where a specific encoding procedure (splitting data and encoding them separately) is enforced. In this respect, we expect CAS to outperform CPS. On the other hand, when full CSI on both the main and eavesdropper channel is available at the transmitter, the two schemes achieve the same performance \cite{Li06,Jorswieck2008a}. Then, it is interesting to see the performance gap when Alice has full CSI on the channel to Bob, and only partial CSI on the channel to Eve. 
%
%Note also that CPS can use up to $K$ different wiretap channel codes, each designed for the corresponding channel conditions, thus increasing the complexity of the transceiver and requiring additional signaling. CAS, on the contrary, requires the choice of a single code, thus reducing both complexity and signaling.
%
Moreover, when a finite message of fixed size is considered, the code length in CAS is larger than in CPS, thus providing an advantage for CAS. %On the other hand, for CPS each sub-message code (rate) can be optimized on the basis of the corresponding channel gain between Alice and Bob, thus providing more flexibility than CAS. 
On the other hand, choosing a CPS scheme yields a parallel implementation of encoding and decoding, allowing the use of solutions devised for AWGN channels.
Because of the advantages and limits of each scheme, it is difficult to establish the superiority of one solution over the other in absolute terms.
}

\section{Secrecy Performance Bounds with Ideal Codes}
\label{sec:three}

We suppose that Alice knows the channel with respect to Bob before transmission, while the Alice-Eve channel is known only in statistical terms\footnote{In Appendix \ref{app:BobChanStat} we consider the case where also the Alice-Bob channel is known only statistically.}. This is a very realistic assumption, since in most of the practical cases Alice does not know the eavesdropper precise location. On the other hand, the Alice-Bob channel state can be learned by conventional channel estimation techniques. Note that we assume that each channel is constant for the whole duration of the transmission, thus allowing for its estimation. %A relevant practical example is OFDM, where the assumption of independent and identically distributed gains over the parallel channels can be met when the subcarriers used for secret transmission are sufficiently spaced apart in frequency (e.g., by leaving the other subcarriers available for the transmission of non-confidential messages), as such subcarrier gains can be obtained via a fast Fourier transform (FFT) of the same size of the vector containing the independent time-domain channel gains.
\myhlF{A relevant practical example is OFDM, where the assumption of \ac{i.i.d.} gains over the parallel channels can be met by considering ideal interleaving across sub-carriers separated by significantly more than the channel coherence bandwidth \cite[p. 101]{Tse}.}

Due to partial \ac{CSI} by Alice on her channel to Eve and to the fact that a finite number $K$ of fading states are spanned by each transmission, we cannot ensure \myhlN{strong} secrecy. Instead, we impose that the probability that Eve gets \myhlN{non vanishing} information on the secret message ({\em strong secrecy outage probability}) is below a given threshold $\epsilon$. 

%\marker{Qui sopra il rev.~1 chiedeva una ''formal definition'' di secrecy, ma questo ci richiedrebbe di} \marker{introdurre simboli per il messaggio segreto e per il segnale trasmesso che attualmente non abbiamo}

In this section we focus on ideal codes, i.e., codes with infinite code length and Gaussian codewords.
We aim at allocating power over the channels for the two coding schemes in order to maximize the secrecy rate while ensuring the target outage probability. 

\paragraph*{Constrained Secrecy Rate Maximization Problem for CPS} 
\myhl{
Let us define the vector of secrecy rates $\bm{R} = [R_1, \ldots, R_K]$ \myhlN{and let 
\begin{equation}
\begin{split}
p_s&(\bm{P},\bm{R};\bm{H}) = \\ 
& \P{\cup_{k=1}^K \{\log (1 + H_k P_k ) - \log(1 +  G_k P_k) \leq R_k\}}
\end{split}
\label{eq:defoutcps}
\end{equation}
be the secrecy outage probability, i.e., the probability that any of the $K$ channels is in secrecy outage. In (\ref{eq:defoutcps}), $\log(\cdot)$ denotes the base-2 logarithm and $\P{\cdot}$ the probability operator, in this case with respect to the random variable $G_k$, while $H_k$ is known.} The following constraints must be satisfied on $\bm{R}$ and $\bm{P}$ 
\begin{subequations}
\label{const1}
\begin{gather}
\myhlF{p_s(\bm{P}, \bm{R}; \bm{H})} \leq \epsilon\,,  \label{Pcon}\\
\frac1K\sum_{k=1}^K P_k \leq P_{\rm max}\,, \label{ac1} \\
P_k \geq 0\,, \quad k=1,2, \ldots, K\,, \label{ac2} \\
R_k \geq 0\,, \quad k=1,2, \ldots, K\,.  \label{ac3} 
\end{gather}
\end{subequations}
Constraint \eqref{Pcon} sets the maximum allowed secrecy outage probability to $\epsilon$; constraint \eqref{ac1} imposes a bound on the average transmit power to $P_{\rm max}$, while \eqref{ac2} and \eqref{ac3} ensure that the resulting powers and rates are non-negative. Now, we aim at finding the maximum average outage secrecy rate {\cite{Zhou11}} that can be achieved, as the solution of the following problem 
\begin{equation}
\max_{\{P_k, R_k\}} \sum_{k =1}^K R_k\,,
\label{problem}
\end{equation}
subject to (\ref{const1}).
}

\paragraph*{Constrained Secrecy Rate Maximization Problem for CAS}
\myhlN{
For CAS, coding is performed by Alice across the $K$ sub-messages. For a given realization of Bob's channel values $(H_1,\ldots,H_K)$, the secrecy outage probability with respect to the random (and unknown to Alice) gains in Eve's channel can be written as {\cite{Gopala08}}
\begin{equation}
\begin{split}
\myhlF{p_s&(\bm{P}, \erreuno;\bm{H})} = \\
&\P{ \sum_{k=1}^K \log(1 + H_k P_k) - \sum_{k=1}^K \log(1 + G_k P_k)  \leq \erreuno}\,.
\end{split}
\label{eq:psCond}
\end{equation}
}
\myhl{Constraint (\ref{Pcon}) becomes
\begin{subequations}
\label{conCAS}
\begin{equation}
\label{Pcon2}
\myhlF{p_s(\bm{P}, \erreuno;\bm{H})} \leq \epsilon\,,
\end{equation}
and (\ref{ac3}) becomes
\begin{equation}
\label{ac32}
\erreuno \geq 0\,,
\end{equation}
\end{subequations}
while the other constraints remain unchanged. The maximization problem becomes
\begin{equation}
\max_{\{P_k, \erreuno\}} \erreuno\,,
\label{problemCAS}
\end{equation}
subject to (\ref{ac1})-(\ref{ac2}) and (\ref{conCAS}).

Note that the constrained maximization problems differ from the conventional bit and power loading for insecure transmission due to the presence of the security constraint. Therefore, the waterfilling solution is not optimal in this case, as will be confirmed by numerical results in Section \ref{sec:five}.}

\subsection{Coding Per Sub-Message}
\label{par:BlockFading}

With CPS each sub-message is encoded independently of the others, with a target secrecy rate $R_k$. Secrecy outage is experienced when at least one of the $K$ sub-messages transmitted over the different channels is in outage. Therefore, the secrecy outage probability in (\ref{eq:defoutcps}) is given by
\begin{equation}
\myhlF{p_s(\bm{P},\bm{R};\bm{H})} = 1 - \prod_{k=1}^K (1- p_k),
\label{eq:outcps}
\end{equation}
where $p_k$ is the secrecy outage probability for sub-message $k$, given that the corresponding realization of $H_k$ is known, i.e.,
\begin{equation}
\begin{split}
&p_k =   \P{\log (1 + H_k P_k ) - \log(1 +  G_k P_k) \leq R_k} \\
& =  \begin{cases}
 1, & R_k  > \log(1+ H_k P_k) \\
1 - F_G\left( \frac{1+H_k P_k}{P_k 2^{R_k}} - \frac{1}{P_k} \right), & \mbox{otherwise}, 
\end{cases}
\end{split} \label{pkCPS}
\end{equation}
where $F_G(x)$ denotes the \ac{CDF} of the eavesdropper power gain $G_k$ over the $k$-th channel, which is the same for all channels. 

\myhlN{If for some $k$ we have $R_k  > \log(1+ H_k P_k)$, then the CPS system is always in outage. Otherwise, by the assumption of \ac{i.i.d.} Rayleigh channel gains, with simple algebra we obtain 
\begin{equation}
\begin{split}
p_s(\bm{P},\bm{R};&\bm{H}) = 1 - \\
&\prod_{k=1}^K \left\{ 1  - \exp  \left[ -\frac{1}{\alpha_{\rm E}} \left(  \frac{1+H_k P_k}{P_k 2^{R_k}} - \frac{1}{P_k}  \right) \right]   \right\}.
\end{split}
\label{ps_Ray}
\end{equation}}
Note that the secrecy outage probability for each sub-message $k$ is a function of both the power $P_k$ and the target secrecy rate $R_k$. Therefore, the rate maximization problem ({\protect\ref{problem}}) cannot be formulated as a special instance of the compound parallel Gaussian wiretap channel {\cite{LiuISIT08}}, since the allocation of the target secrecy rates $R_k$ adds $K$ variables to the rate maximization problem. In fact, constraint ({\protect\ref{Pcon}}) can be met by different rate $K$-tuples. The secrecy rates $R_k$ are related to the transmit power. In particular, if we restrict the constraints (\ref{ac2})--(\ref{ac3}) to hold without equality, i.e.,
\begin{equation}
P_k > 0\,, \quad R_k > 0
\label{ac2-3strict}
\end{equation}
and denote by $\bar{p}_k$ the target secrecy outage probability for each sub-message, i.e.,
\begin{equation}
 \exp  \left[ -\frac{1}{\alpha_{\rm E}} \left(  \frac{1+H_k P_k}{P_k 2^{R_k}} - \frac{1}{P_k}  \right) \right] = \bar{p}_k\,,
\label{poutset}
\end{equation}
we obtain the following result. 

\begin{teo}
\label{th:RateAllocation}
For a given $K$-tuple of outage probabilities $\bar{\bm{p}} = (\bar p_1,\ldots,\bar{p}_K)$ and $\nu > 0$, let us define $\bar{u}_k = - \alpha_{\rm E}\ln \bar{p}_k$. If $\nu  -   H_k +  \bar{u}_k < 0$, $\forall k$, then the power allocation
\begin{equation}
\begin{split}
P_k = & P_k\opt =  \frac{-\nu  (\bar{u}_k + H_k)}{2\bar{u}_k\nu  H_k} +  \\
 & \frac{\sqrt{\left[\nu  (\bar{u}_k + H_k)\right]^2 - 4\bar{u}_k\nu  H_k (\nu  -   H_k +  \bar{u}_k)}}{2\bar{u}_k\nu  H_k}\,
\end{split}
\label{potott}
\end{equation}
maximizes the sum-rate \eqref{problem} under the constraint \eqref{ac2-3strict}. %--\eqref{poutset}.

Therefore, if  $\nu$ is such that \eqref{ac1} is satisfied with equality, $\{ P_k\opt \}$ is also the power allocation  that solves the maximization problem \eqref{problem}, under constraints \eqref{Pcon}, \eqref{ac1} and \eqref{ac2-3strict}, with secrecy outage probability as in \eqref{ps_Ray}. The corresponding secrecy rate of the message is 
\begin{equation}
R\sub s = \sum_{k = 1}^K R_k = \sum_{k=1}^K \log \frac{1+H_k P_k}{\bar{u}_k P_k + 1  }\,.
\label{secrate}
\end{equation}
\end{teo}

\begin{IEEEproof}
See Appendix~\ref{par:ProofRate}.
\end{IEEEproof}

From Theorem \ref{th:RateAllocation} we conclude that the maximum secrecy rate ensuring a secrecy outage probability not greater than $\epsilon$ is obtained by solving
%\begin{subequations}
\begin{equation}
\max_{\bar{\bm{p}}, \nu} \sum_{k=1}^K \log \frac{1+H_k P_k}{\bar{u}_k P_k + 1  } \label{problem2}
\end{equation}
subject to $1 - \prod_{k=1}^K (1- \bar{p}_k) \leq \epsilon$, (\ref{potott}), (\ref{ac1}), and (\ref{ac2-3strict}). The solution of these problems requires numerical methods. Note however that Theorem \ref{th:RateAllocation} allows a strong reduction in the number of unknowns: from $2K$ in the original problem formulation (\ref{problem}) to $(K+1)$ in the formulation (\ref{problem2}).

\subsection{Coding Across Sub-Messages}
\label{subsec:CAS}

\myhlFC{
We start from (\ref{eq:psCond}), which reflects the specific encoding (and decoding) structure of CAS, and in this respect it is different from (\ref{eq:defoutcps}), valid for CPS.
}

Let 
\begin{equation}
\myhlFC{\Phi(\bm{P}, \erreuno;\bm{H})} = 2^{ \left[ \sum_{k=1}^K\log (1 +H_k P_k ) - \erreuno \right]}
\label{defPhi}
\end{equation}
and let us define
\begin{equation}
\myzeta(\bm{P}) = \prod_{k=1}^K e^{\frac{1}{{P_k \alpha_{\rm E}}}}\,, \quad \phi(\bm{P}) = \left[\prod_{k=1}^K \frac{1}{{P_k \alpha_{\rm E}}} \right]^{-1}.
\end{equation} 
As derived in Appendix  \ref{prooffinal}, the secrecy outage probability (\ref{eq:psCond}) can be written as 
\begin{equation}
\begin{split}
\myhlF{p_s& (\bm{P}, \erreuno;\bm{H})} =  1 - \frac{\myzeta(\bm{P})}{\phi(\bm{P})} \times \\ 
& \left\{ \myhlFC{\Phi(\bm{P}, \erreuno;\bm{H})} \mathcal G\left(\frac{\myhlFC{\Phi(\bm{P}, \erreuno;\bm{H})}}{\phi(\bm{P})}\right() - \mathcal G \left(\frac{1}{\phi(\bm{P})}\right) \right\}\,,
\end{split}
\label{psfinal}
\end{equation}
where
\begin{equation}
\begin{split}
\mathcal G&(a) =  \\ 
&\mathcal H_{1,K+1}^{K,1}  \left[a\left|\begin{array}{c} \{(0, 1, 0)\} \\ \{\{(0, 1,(P_k \alpha_{\rm E})^{-1}\}_{k = 1, \ldots, K},\, (-1, 1, 0)\} \end{array} \right.\right]
\end{split}
\end{equation}
and $\mathcal H[\cdot]$ is the generalized Fox H-function, whose definition is recalled in  Appendix \ref{prooffinal}.

\begin{comment}
Unfortunately, no closed form expression of the outage probability \eqref{eq:psCond} is available. Therefore we resort to an approximation {\cite{Wu-apr10}}, modeling $\log(1 + G_k P_k)$ as a Gaussian variable with mean
\begin{equation}
\mu(P_k) = \log(e) e^{1/(\alpha_{\rm E} P_k)}{\rm E}_1\left(\frac{1}{\alpha_{\rm E} P_k}\right)\,,
\end{equation}
and variance 
\begin{equation}
\sigma^2(P_k) = \frac{2}{\alpha_{\rm E} P_k}\log(e)^2  e^{1/(\alpha_{\rm E} P_k)} G_{3,4}^{4,0}\left(1/(\alpha_{\rm E} P_k)|_{0,-1,-1,-1}^{0,0,0} \right) - \mu^2(\alpha_{\rm E} P_k)
\end{equation}
where ${\rm E}_1(x) = \int_1^\infty t^{-1} e^{-xt} {\rm d}t$ and $G_{p,q}^{(m,n)}(x|_{b_1, \ldots, b_q}^{a_1, \ldots, a_p})$ is the Meijer G-function.

Under the Gaussian approximation, (\ref{Pcon}) can be rewritten as 
\begin{equation}
{\rm Q}\left(\frac{\sum_{k=1}^K \left[\log(1 + H_k P_k) - \mu(P_k)\right] - K\erreuno}{\sqrt{\sum_{k=1}^K \sigma^2(P_k)}}\right) \leq \epsilon,
\end{equation}
where ${\rm Q}(\cdot)$ is the tail probability of a unit variance normal random variable, or equivalently
\begin{equation}
0 \le \erreuno \leq \frac1K\sum_{k=1}^K \left[\log(1 + H_k P_k) - \mu(P_k) \right]- \frac{{\rm Q}^{-1}(\epsilon)}K \sqrt{\sum_{k=1}^K \sigma^2(P_k)}\,.
\label{eq:rateCA}
\end{equation}
\end{comment}

\myhl{
Then (\ref{Pcon2}) can be rewritten as 
\begin{equation}
0 \le \erreuno \leq \myhlF{p_s^{-1}(\bm{P}, \epsilon;\bm{H})}\,,
\label{eq:rateCA}
\end{equation}
where $\myhlF{p_s^{-1}(\bm{P}, \epsilon;\bm{H})}$ is the inverse of (\ref{psfinal}) with respect to $\erreuno$.

When the outage secrecy rate is maximized, $\erreuno$ equals the right hand side (r.h.s.) in \eqref{eq:rateCA}; therefore, we can remove $\erreuno$ from the optimization variables and the maximum outage secrecy rate problem (\ref{problemCAS}) can be rewritten as % \marker{qui ho tolto l'$1/K$ che mi sembra non coerente con la (3) e la (7)}
\begin{equation}
\label{eq:MaxrateCA}
\max_{\bm{P}} \myhlF{p_s^{-1}(\bm{P}, \epsilon;\bm{H})}\,,
\end{equation}
subject to (\ref{ac1}) and (\ref{ac2}). This problem cannot be solved in closed form and we must resort to numerical methods. Examples will be given in Section \ref{sec:five}.
}

\section{Secrecy Performance Bounds With Practical Codes}
\label{sec:four}

\myhlN{
The analysis in Section \ref{sec:three} relies on the use of ideal codes, thus providing an upper bound on the performance reachable by using practical forward error correcting (FEC) codes.
Indeed, when long codewords can be used, practical codes (e.g., LDPC codes) may well approximate asymptotic
performance achieving results close to capacity.  In this context, the transmission rate to Bob that can be obtained with the power allocation $\bm{P}$ corresponds to the code rate, 
while its difference with the outage secrecy capacity provides the rate of the random message to be used in random binning in order to obtain the target secrecy outage probability.
However, when limits on the length of the codeword are relevant, due to delay constraints or channel coherence time concerns, other approaches should
be considered in the code design.}
Hence, in order to better assess the performance of finite length practical codes, we gradually introduce the characteristics of a real transmission:
\begin{itemize}
\item {\bf Discrete (finite) constellations:} when finite and discrete constellations are used, in the secrecy capacity expression we should consider the constellation-constrained mutual information.
\item {\bf Deterministic encoding:} when FEC is used without probabilistic encoding (e.g., random binning) that is typical of wiretap codes, we focus on the amount of information per channel use that remains unknown to Eve, \myhlN{with probability at least $1-\epsilon$; we thus introduce the $\epsilon$-outage equivocation rate, extending the notion given in  \cite{WongWong2011} for \ac{AWGN} channels}.
\item {\bf Finite length codes:} with finite length codes, both Bob and Eve are prone to errors and secrecy cannot be assessed by Eve's equivocation only. The metric that suitably summarizes the error probabilities of the two agents is the security gap (used for the \ac{AWGN} channel in \cite{Klinc2011, Baldi2010, Baldi2011, Baldi2012}), \myhlFC{as will be defined in Section \ref{subsec:ErrorRate}.}
%, defined as the ratio of the minimum \ac{SNR} required to ensure both a low error probability for Bob and a high error probability to Eve.
\end{itemize}

\subsection{Finite Constellation and Deterministic Encoding}
\label{subsec:EquivocationRate}

\myhl{
When a finite constellation is considered (still with wiretap coding), let \myhlF{$C(\gamma)$ be} the \myhlN{mutual information rate} of a Gaussian channel {with a fixed (e.g., uniform) distribution} as a (monotonically increasing) function of the \ac{SNR} $\gamma$. The expression of $C(\gamma)$ depends on the adopted input constellation.

In this case, for CPS, (\ref{pkCPS}) becomes
\begin{equation}
p_k = \P{C(H_kP_k) - C(G_kP_k) \leq R_k}\,,
\end{equation}
thus providing for Rayleigh fading channels
\begin{equation}
\label{e:IeCPSbnd}
\begin{split}
\myhlF{p_s(\bm{P}, &\bm{R}; \bm{H})} =  1- \\
&\prod_{k=1}^K \left\{1-\exp\left(-\frac{C^{-1}[C(H_kP_k) - R_k]}{P_k\alpha_{\rm E}}\right)\right\}\,.
\end{split}
\end{equation}
Similarly, for CAS, (\ref{eq:psCond}) becomes
\begin{equation*}
\begin{split}
\myhlFC{p_s& (\bm{P}, R; \bm{H})}  =   \P{\sum_{k=1}^{K} [C(H_kP_k) - C(G_kP_k)] \leq \erreuno} \\
&  \leq  1-\prod_{k=1}^K \left\{1-\exp\left(-\frac{C^{-1}[(\sum_kC(H_kP_k) - R)/K]}{P_k\alpha_{\rm E}}\right)\right\} ,
\end{split}
\label{e:IeCASbnd}
\end{equation*}
where the last upper bound is obtained by assuming Rayleigh fading channels and by observing that 
\begin{equation}
\sum_{k=1}^{K} C(G_kP_k) \leq K \max_{k} C(G_kP_k).
\end{equation}
%thus providing the right size of (\ref{e:IeCPSbnd}). In other word, (\ref{e:IeCPSbnd}) is both the exact secrecy outage probability for CPS and a bound for that of CAS. Hence, we conclude that CAS has in general a lower constrained capacity than CPS. 
The outage secrecy rate obtained under constellation constrained transmission is denoted as $C_{\rm s}^{(\epsilon)}$.

Consider now a deterministic encoding without wiretap coding features, with a fixed code rate $R\sub c$. In {\cite{WongWong2011}} the level of confidentiality obtained in a coded transmission over an \ac{AWGN} channel is evaluated through its equivocation rate, that is, the difference between the code rate and the information rate at the eavesdropper. The equivocation rate is an indicator of the residual uncertainty of the eavesdropper on the transmitted message. We extend the notion of equivocation rate to the considered scenario through an outage formulation, and derive lower bounds for both CPS and CAS. The code rate $R\sub c$ and allocated powers $\setg{P_k}$ are assumed to satisfy the reliability conditions
\begin{subequations}
\begin{equation}
R\sub c \leq C(H_kP_k)\;\;\;\;\; \forall k \quad \mbox{ for CPS,}
\end{equation}
\begin{equation}
R\sub c \leq \myhlN{\frac1K} \sum_{k=1}^K C(H_kP_k) \quad \mbox{ for CAS.}
\end{equation}
\end{subequations}
Since all information bits are intended for confidential transmission, now $R\sub c$ plays the role of mutual information between Alice and Bob. In particular, the secrecy outage probabilities have the expressions (\ref{e:IeCPSbnd}) and (\ref{e:IeCASbnd}), where $C(H_kP_k)$ is replaced by $R\sub c$. Correspondingly, the $\epsilon$-outage equivocation rate is the maximum value of $\sum_k R_k$ or $R$ such that the outage probability constraint is satisfied.
}

\subsection{Finite Length Codes}
\label{subsec:ErrorRate}
\myhl{
When codes of finite length are considered, we cannot use the secrecy capacity or the equivocation rate to assess their performance. Instead we have to take into account the non-vanishing error probability incurred by these codes. Let us denote by $p^{\rm B}$ and $p^{\rm E}$ the decoding error rate on the entire message received by Bob and Eve, respectively. Given two arbitrarily small threshold values, $\rho$ and $\eta$, the transmission can be considered reliable and secure if the following two conditions are satisfied\footnote{Note that condition (\ref{condber2}) refers to the decoding error probability. For messages that are not perfectly source-coded (i.e., are not at maximum entropy), non-systematic codes must be used to increase secrecy \cite{Klinc2011,Baldi2012}.\par Note that, assuming that the secret message is uniformly distributed, perfect secrecy requires that (\ref{condber2}) is satisfied with $\eta = 2^{-NR\sub s}$ for any decoding strategy, where $N$ is the codeword length. Such condition is also sufficient if it holds in particular for the optimal strategy, that is \ac{ML} decoding. Similarly, for the rate $R\sub s$ to be achievable under an information theoretic secrecy criterion of variational distance (e.g., criterion (2) in \cite{BlochLaneman08}) it is necessary and sufficient that there exists a sequence of codes with lengths $N \in \mathbb{N}$, such that (\ref{condber2}) is satisfied under ML decoding for each $N$ with $\eta \sim 2^{-NR\sub s}$ as $N\rightarrow\infty$, while at the same time $\rho\rightarrow 0$ in \myhlFC{(\ref{condber1})}.} \myhlFC{\cite{Klinc2011}, \cite{Baldi2012}}:

\begin{subequations}
\label{condber}
\begin{equation}
p^{\rm B} \le \rho\,, 
\label{condber1}\end{equation}
\begin{equation}
p^{\rm E} \ge 1 - \eta\,.
\label{condber2}
\end{equation}
\end{subequations}

\myhlF{
The condition on $p^{\rm B}$ for CPS can be translated into a condition on the \ac{CER} $p_k^{\rm B}$ on each sub-message $k$. We first observe that 
\begin{equation}
p^{\rm B} = 1 - \prod_{k=1}^K (1 - p_k^{\rm B})\,.
\end{equation}
Although in general the maximum secrecy outage rate is achieved with different values of $p_k^{\rm B}$ for each channel, here we focus on the case of equal error probabilities for each channel, so that (\ref{condber1}) becomes
\begin{equation}
p_k^{\rm B} \leq 1 - \sqrt[K]{1 - \rho} \,.
\end{equation}
For CAS, instead, the condition on $p^{\rm B}$ directly translates into a condition on the \ac{CER}, since $p^{\rm B}$ coincides with the \ac{CER} in this case. Therefore, we can fix a threshold $\delta$ on the \ac{CER} for the two schemes as follows
\begin{equation}
\left\{
\begin{array}{ll}
p_k^{\rm B} \le \delta = 1 - \sqrt[K]{1 - \rho}\;, & \mbox{for CPS},\\
p^{\rm B} \le \delta = \rho\;, & \mbox{for CAS}.
\end{array}
\right.
\label{condber1ter}
\end{equation}
}
%\hl{fine sostituzione. Passerei tutti i $\delta$ a $\rho$ tranne nella right side della} (\ref{dg2}) \hl{ dove mettiamo  $1 - \sqrt[K]{1 - \rho}$. Occhio all'Appendice C. }

%In the CPS case the above condition is sufficient yet not necessary for \eqref{condber1}. However, it allows us to set the same threshold value on each channel, regardless of the error rate on the other channels. Hence, we adopt this choice for the sake of simplicity.

On the contrary, we impose that the \ac{CER} for Eve always equals or overcomes $1 - \eta$.
It is important that this occurs even with CPS, on each channel, since otherwise Eve, though not being able to decode the whole message, could successfully discover some part of it.
}

Note that condition \eqref{condber1} can be met through a suitable power allocation,
since Bob's channels are known.
Condition \eqref{condber2}, instead, can only be met statistically, that is, by tolerating some outage probability,
% \marker{qui non mi piace riusare ``secrecy outage'' che abbiamo definito nella Section III}
since only a statistical description of Eve's channels is available.
The case in which Bob's channels are also known only in statistical
terms is studied in Appendix \ref{app:BobChanStat}.

\myhlF{
We indicate by $\gamma_\delta^{\rm B}$ the minimum \ac{SNR} on each channel that ensures condition \eqref{condber1ter}, that is
\begin{equation}
 \gamma_\delta^{\rm B}(k) = \min\left\{\gamma\in\mathbb R\,:\,\P{E_k^{\rm B}|P_kH_k = \gamma} \leq \delta \right\}
 \label{dg2}
 \end{equation}
where $E_k^{\rm B}$ denotes Bob's decoding error event on channel $k$ for CPS, and
\begin{equation}
\begin{split}
\gamma_\delta^{\rm B} = & \min\left\{\gamma\in\mathbb R\,:\,\right. \\
& \left. \P{E^{\rm B}|P_1H_1 = \gamma, \ldots,P_KH_K = \gamma} \leq \delta\right\}
\end{split}
\label{dg1}
\end{equation}
with $E^{\rm B}$ denoting Bob's decoding error event for CAS.} 
On the other hand, since we assume that Alice does not know Eve's channels, we consider an outage approach for the definition of the security gap. In fact, $p^{\rm E}$ is a random variable, whose distribution, under the \ac{i.i.d.} Rayleigh assumption, only depends on the average \ac{SNR} of Eve's channels\myhlFC{, defined as}  
\begin{equation}
\bar{\gamma}^{\rm E} = \frac{1}{K} \sum_k \alpha_{\rm E} P_k.
\label{eq:GammaBarE}
\end{equation}

We are interested in finding the maximum value of $\bar{\gamma}^{\rm E}$, denoted by $\bar{\gamma}_{\max}^{\rm E}$,
for which the probability that $p^{\rm E} < 1-\eta$ is not greater than $\myzeta$.
In fact, $\bar{\gamma}_{\max}^{\rm E}$ represents the maximum average \ac{SNR} over Eve's channel which is acceptable to meet \eqref{condber2} under the outage constraint.
We have
\begin{equation}
\bar{\gamma}_{\max}^{\rm E} = \frac{1}{K} \sum_k P_k  \max \{\alpha_{\rm E}: \P{p^{\rm E} < 1-\eta} \le \myzeta\}\,.
\label{ae1}
\end{equation}
%\myhl{
Aiming to extend the original definition of security gap given for the \ac{AWGN} channel in {\cite{Klinc2009a}} to the considered scenario,
we define the $\myzeta$-outage security gap as 
\myhlF{
\begin{equation}
S_\myzeta = \left\{
\begin{array}{ll}
\frac{\sum_{k =1}^K\gamma_\delta^{\rm B}(k)}{K \,\bar{\gamma}_{\max}^{\rm E}} & \mbox{for CPS},\\
\frac{\gamma_\delta^{\rm B}}{\bar{\gamma}_{\max}^{\rm E}} & \mbox{for CAS}.  
\end{array}
\right.
\label{Sgzeta}
\end{equation}
}
%}
\myhlFC{It has to be observed that the security gap defined by \eqref{Sgzeta} is computed on the basis of the codeword error rate,
and does not depend on the bit error rate of the secret message.
This allows us to define a target which does not depend on the secret message rate, and to study the achievable equivocation rate.
Then, some nested coding approach \cite{Harrison2013} should be used to achieve a secret message rate that approaches the achievable equivocation rate.
Nested codes can also be obtained through the scrambling-based non-systematic encoding approach proposed in \cite{Baldi2010, Baldi2011, Baldi2012},
by using a code with length $N$ and dimension $N\sub d = R\sub c N$, and $N\sub s \le N\sub d$ bits for the secret message.
The $N\sub d - N\sub s$ remaining information bits are randomly generated, therefore each secret message is randomly associated to $2^{N\sub d-N\sub s}$ codewords.
This, however, is out of the scope of this paper.
}

In the following, we derive bounds on the $\myzeta$-outage security gaps for  CPS and CAS,  and discuss the optimization of the corresponding power allocations.

\subsection{Computation of $\bar{\gamma}_{\max}^{\rm E}$ and $\gamma_{\delta}^{\rm B}$}
\label{subsec:GammaMaxE}

The parameters required for the computation of the security gap \eqref{Sgzeta} are now derived.
%For explicative purposes, we consider first CPS and then CAS.
\paragraph*{Computation of $\bar{\gamma}_{\max}^{\rm E}$ for CPS} When coding is applied separately on each sub-message, condition \eqref{condber2} must hold on each channel. 

The eavesdropper outage probability becomes
\begin{IEEEeqnarray}{rCl}
p^{\rm E} &=& 1 - \prod_{k=1}^K \left(1 - \P{p_k^{\rm E} < 1-\eta}\right) \nonumber \\
& = & 1- \prod_{k=1}^K (1 - \P{P_kG_k > \gamma_\eta^{\rm E}(k)})\,,
\label{eq:PCPS}
\end{IEEEeqnarray}
where $p_k^{\rm E}$ is Eve's \ac{CER} on the $k$-th channel, and $\gamma_\eta^{\rm E}(k)$ is the maximum \ac{SNR} on the same channel which ensures $p_k^{\rm E} \geq 1 - \eta$, that is
$$ \gamma_\eta^{\rm E}(k) = \max\left\{\gamma\,:\,\P{E_k^{\rm E}|P_kG_k = \gamma}\geq 1-\eta\right\}$$
with $E_k^{\rm E}$ denoting Eve's decoding error event on channel $k$. 
In Appendix \ref{appeta} we derive a lower bound on the value of $\gamma_\eta^{\rm E}(k)$ which permits us to estimate the best performance achievable by Eve. From the Rayleigh distribution assumption we have 
\begin{equation}
\P{P_kG_k > \gamma_\eta^{\rm E}(k)} = \exp \left( - \frac{\gamma_\eta^{\rm E}(k)}{P_k \alpha_{\rm E}} \right) \label{Pray}
\end{equation}
and hence \eqref{eq:PCPS} becomes
\begin{equation}
p^{\rm E} = 1- \prod_{k=1}^K \left[1 - \exp \left( - \frac{\gamma_\eta^{\rm E}(k)}{P_k \alpha_{\rm E}} \right)\right]\,.
\label{PMgee}
\end{equation}
By exploiting these expressions, and the knowledge of the transmission (and reception) technique, we can compute
$\bar{\gamma}_{\max}^{\rm E}$ for which condition \eqref{condber2} is satisfied for a given power allocation.

\paragraph*{Computation of $\bar{\gamma}_{\max}^{\rm E}$ for CAS}  
\myhl{
When CAS is implemented, no closed form expression exists for Eve's \ac{CER}; hence we resort to a lower bound. In particular, since the \ac{CER} is a non-increasing function of the \ac{SNR} on each channel, we have
\begin{equation}
\begin{split}
\P{E^{\rm E}|P_1G_1 = \gamma_1,\ldots,P_KG_K = \gamma_K} \geq \\
\P{E^{\rm E}|P_1G_1 = \gamma_M,\ldots,P_KG_K = \gamma_M}
\end{split}
\label{lbPE}
\end{equation}
where $M = \arg\max_k \gamma_k$ is the index of the channel with maximum \ac{SNR}. Hence, a sufficient condition for \eqref{condber2} is that $\max_k P_kG_k \leq \min_k \gamma_\eta^{\rm E}(k)$, thus we have 
\begin{equation}
\P{p^{\rm E} < 1-\eta} \leq \P{\max_k\{P_kG_k\} > \min_k \gamma_\eta^{\rm E}(k)}
\label{pre-ae2}
\end{equation}
and we can replace \eqref{ae1} with
\begin{equation}
\begin{split}
\bar{\gamma}_{\max}^{\rm E} = & \frac{1}{K} \sum_k P_k  \max \{\alpha_{\rm E}: \\ 
& \P{\max_k\{P_kG_k\} > \min_k \gamma_\eta^{\rm E}(k)} \le \myzeta\}\,.
\end{split}
\label{ae2}
\end{equation}
The probability on the r.h.s. of \eqref{pre-ae2} can be calculated as
\begin{equation}
\begin{split}
& \P{\max_k \{P_kG_k\} > \min_k \gamma_\eta^{\rm E}(k)} = \\ &\P{\bigcup_{k=1}^K \{P_kG_k > \min_k \gamma_\eta^{\rm E}(k)\}} = \\
& 1- \prod_{k=1}^K \left[1 - \exp \left( - \frac{\min_k \gamma_\eta^{\rm E}(k)}{P_k \alpha_{\rm E}} \right)\right]\,.
\end{split}
\label{PMgeeCAS}
\end{equation}

Therefore, we obtain an expression similar to \eqref{PMgee}, though in this case it results from the use of the lower bound \eqref{lbPE}, while in the CPS case it is given by an exact derivation.
For the special case in which $\gamma_\eta^{\rm E}(1) = \gamma_\eta^{\rm E}(2) = \ldots = \gamma_\eta^{\rm E}(K) = \gamma_\eta^{\rm E}$, these two expressions coincide; 
so, we can use the same formula to model both the CPS and CAS scenarios.
}

\paragraph*{Computation of $\gamma_{\delta}^{\rm B}$}

In order to model Bob's channels, which are supposed to be known, we only need to compute $\gamma_{\delta}^{\rm B}$, that is, the threshold value of the channel gains which allows constraints \eqref{condber1ter} to be satisfied. % with equality.
%The case in which Bob's channels are known only in statistical terms is more involved and, as mentioned, it is addressed in Appendix \ref{app:BobChanStat}.

In this part of the analysis, we refer to \ac{ML} decoding also for Bob, and use the well-known union bound to obtain an upper bound on $p^{\rm B}$.
In fact, in the high \ac{SNR} region, the union bound is known to provide a tight approximation of the performance of \ac{ML} and \ac{ML}-like decoders {\cite{Garello2001}}. % (like, for example, LDPC decoders).

Let us consider a linear block code with codeword length $N$, and let $d_{\min}$ denote the code minimum distance and $A_w$ the number of codewords with weight $w$.
If we focus on a single channel with \ac{SNR} $\gamma$, we have the following bound on the \ac{CER}
\begin{equation}
p^{\rm B}  \le \sum_{w = d_{\min}}^N A_w {\rm Q}\left( \sqrt{2 \gamma w } \right), 
\label{eq:UnionBound}
\end{equation}
where ${\rm Q}(x) = \frac{1}{\sqrt{2\pi}}\int_x^\infty e^{-t^2/2}{\rm d}t$ is the complementary {\ac{CDF}} of the zero-mean, unit-variance Gaussian distribution.
% By considering only a subset of the weight spectrum of the code, we get a truncated union bound approximation.
By considering only the minimum weight codewords, we get the following approximation
\begin{equation}
p^{\rm B}   \simeq A_{d_{\min}} {\rm Q}\left( \sqrt{ 2 \gamma d_{\min} } \right),
\label{eq:TruncUnionBound}
\end{equation}
that provides a very good estimate of \ac{ML} (or \ac{ML}-like) decoding performance for large values of $\gamma$, i.e., small values of \ac{CER}, that are those of interest for Bob.
By using the parameters of the code used in the $k$-th channel, and by equating the r.h.s. of \eqref{eq:TruncUnionBound} to $\delta$ and solving for $\gamma$, we obtain 
$\gamma_\delta^{\rm B}(k)$ under \ac{ML} decoding, as defined in \eqref{dg2}.
\myhl{
Concerning the CAS scenario, according to \eqref{dg1}, we have $\gamma_\delta^{\rm B} = \max_k \gamma_\delta^{\rm B}(k)$.
}

\subsection{Power Allocation}
\label{subsec:SgPowerAlloc}

\myhlFC{Let us} consider fixed secrecy rate transmissions, regardless of the channel state. On the other hand, by varying the power allocation we can alter the decoding reliability at Bob and Eve. Hence, in a parallel to the security rate regions, we see that conditions (\ref{condber}) define regions for power allocation strategies that ensure reliable and secure communications. 

In order to satisfy Bob's reliability condition \eqref{condber1ter}, Alice transmits at minimum power levels
\begin{equation}
P_k = \frac{\gamma_\delta^{\rm B}(k)}{H_k}\,.
\label{Pkcond}
\end{equation}

More precisely, based on \eqref{Pkcond}, Alice finds the optimal power allocation, and checks whether the power
constraint \eqref{ac1} is satisfied or not.
In the former case, transmission occurs. Otherwise, Alice skips the transmission, since the reliability
target cannot be achieved.

\section{Numerical Results}
\label{sec:five}

%\hl{In questa sezione trovo $S_\myzeta$, $S_{\omega, \delta}$ ma la definizione \'e $S_{\myzeta}(\delta, \eta)$. Mi sto confondendo io o la notazione \'e da sistemare?}

On the basis of the theoretical analysis developed in the previous sections, we provide here some examples, under different conditions of the parallel channels.
Since the target of our analysis is not to find an optimal code/allocation strategy for the CPS and CAS schemes, but rather to assess and compare the performance achievable by using a practical code in these configurations, we fix the choice of the code for both CPS and CAS.
Moreover, for CPS, full variable rate coding on each channel could be considered. In this section, however, aiming at practically feasible and simple systems, we use a linear block code with fixed length and rate for all channels.
As a counterpart, this means that CPS performance could be further improved by a proper encoder selection over channels with different bit-loading.
Under this assumption, we have $\gamma_\delta^{\rm B}(1) = \gamma_\delta^{\rm B}(2) = \ldots = \gamma_\delta^{\rm B}(K) = \gamma_\delta^{\rm B}$ and
$\gamma_\eta^{\rm E}(1) = \gamma_\eta^{\rm E}(2) = \ldots = \gamma_\eta^{\rm E}(K) = \gamma_\eta^{\rm E}$.

For both CPS and CAS, we suppose to use binary phase shift keying (BPSK) and a linear block code with length $N = 128$ bits and rate $R_{\rm c} = 1/2$.
We focus on a ($128, 64$) extended BCH (eBCH) code with minimum distance $d_{\min} = 22$. It must be noted that, contrary to the approaches searching for secrecy capacity achieving codes {\cite{Mahdavifar2011, Koyluoglu2012}}, in our analysis the code parameters (rate and length) are fixed. % Our goal, in fact, is not to find a code approaching the theoretical limits but, instead, to provide suitable tools to evaluate and compare the reliability/security performance of any practical code.

The eBCH code, in particular, is good for the chosen length, since soft-decision algorithms can be used for decoding, achieving performance close to that of state-of-the-art LDPC codes. In these conditions, it is realistic to assume that Bob and Eve use the same decoder, and this contributes to keeping the security gap small. On the other hand, \ac{ML}-like decoding becomes intractable for longer codes, while long LDPC codes with soft-decision iterative decoding 
achieve good performance with limited complexity.
The choice of LDPC codes allows Bob to work at a lower \ac{SNR}, but the gap to the theoretical limits increases.
Hence, since we assume that Eve is always able to use the best decoder, the security gap by using long LDPC codes becomes larger than for the case of short eBCH codes with \ac{ML}-like decoding.
\myhlFC{
On the other hand, if we relax this hypothesis, and consider that Eve uses a practical decoder, the resulting security gap becomes smaller. For example, by using long LDPC codes and iterative belief propagation decoding for both Bob and Eve, a security gap reduction of several dBs would result with respect to the case of short eBCH codes with \ac{ML}-like decoding.
}

%Concerning the use of scrambling, as mentioned in Section \ref{sec:four}, we have assumed the availability of a perfect scrambler. In \cite{Baldi2012} we have verified that perfect scrambling can be achieved by using a scrambling matrix with a proper density of symbols $1$; the best scrambling effect is obtained when the density is $0.5$, but a lower density can suffice to approach perfect scrambling. For the considered code, the scrambling matrix has size $64L \times 64L$, where the number $L$ of concatenated codewords is assumed sufficiently large.

\subsection{Coding per Sub-message}
\label{ssCPS}
\subsubsection{Security gap and equivocation rate with eBCH coding}

\begin{figure}[tb]
\begin{centering}
\includegraphics[keepaspectratio, height=70mm]{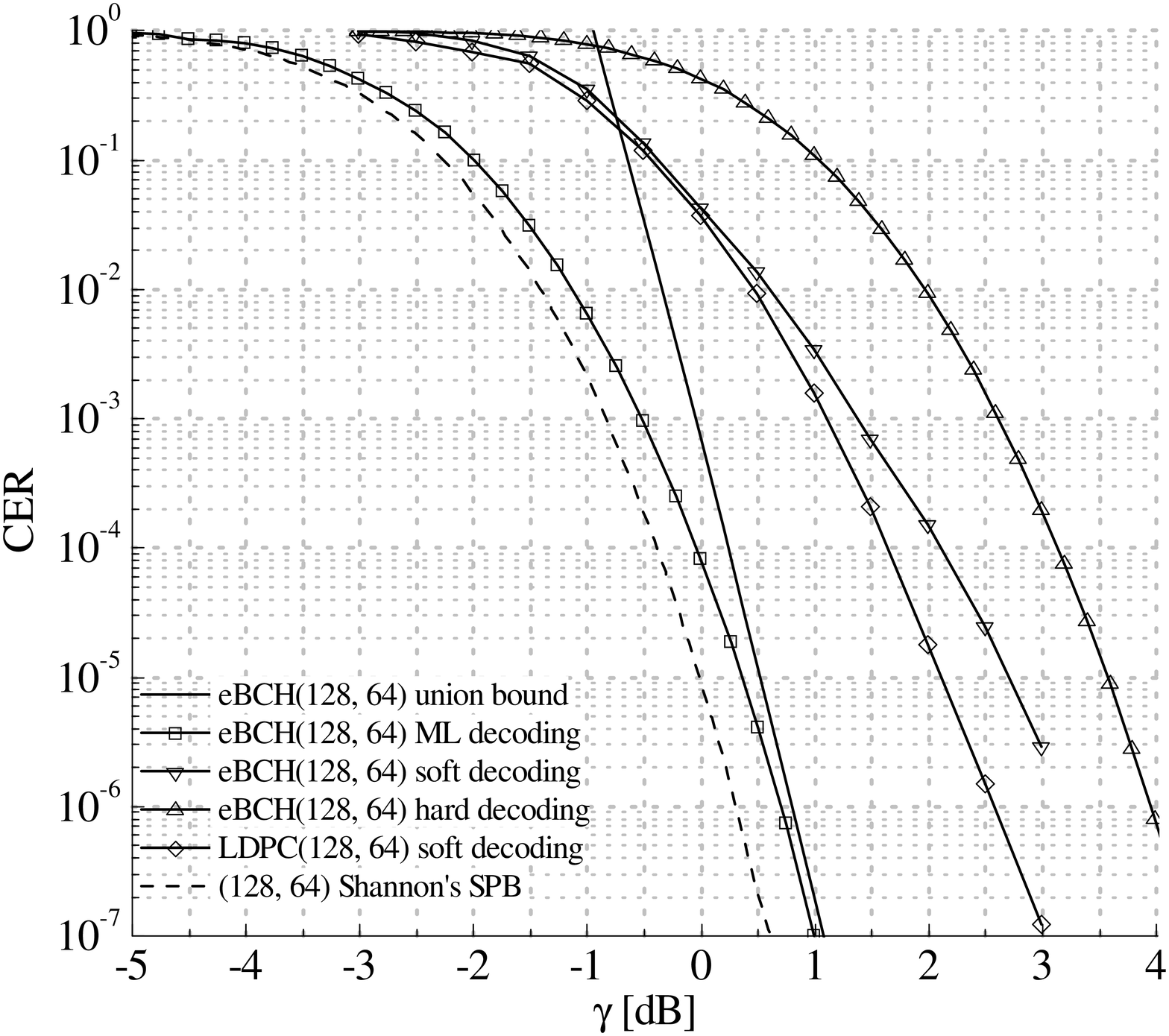}
%\\\marker{io qui nella versione stampata non riesco a vedere le linee "solid", ma sul file le vedo.}{\`E un problema solo mio?} 
\caption{\ac{CER} simulated values and bounds for codes with length $N = 128$ and rate $R_{\rm c} = 1/2$ over a single static channel with \ac{SNR} $\gamma$.}
\label{fig:CodeExample128}
\par\end{centering}
\end{figure}

The  achievable performance for $N = 128$ and rate $R_{\rm c} = 1/2$ is shown in Fig. 3, in terms of \ac{CER} over a single static channel with \ac{AWGN}, as a function of the channel \ac{SNR} $\gamma$. The union bound, computed through \eqref{eq:UnionBound}, and the Shannon's \ac{SPB}, computed as described in Appendix \ref{appeta}, are also shown for the sake of comparison. The performance of the eBCH code under ML decoding is obtained as in {\cite{Valembois2004}}. From the figure we observe that the ML decoding performance is tightly upper bounded by the union bound in the high \ac{SNR} region, and tightly lower bounded by Shannon's \ac{SPB} in the low SNR region. This confirms that the two bounds are well suited to model the performance achievable on each channel by Bob and Eve, respectively. We also report, for the sake of comparison, the performance achieved by using other decoders. Soft-decision decoding of the eBCH code has been implemented by following the approach proposed in {\cite{Jiang2006}}, while hard-decision decoding of the same code has been simply estimated by using the closed form expression for bounded distance decoders {\cite{Wicker1995}}.
We have also included the performance of an LDPC code, with the same length and rate, designed through the progressive edge growth algorithm {\cite{Hu2001PEG}}, and decoded through the logarithmic version of the sum-product algorithm {\cite{Hagenauer1996}}. Shannon's \ac{SPB} provides a lower bound for all the considered schemes, so it actually represents
a reliable and conservative tool for modeling Eve's performance. Instead, when Bob uses other decoders than ML, his performance can be far worse than the union bound. In this case, the security gap must be increased by a suitable margin, which depends on the specific decoding algorithm used by Bob. By focusing on ML decoding for both Bob and Eve, and using the upper and lower bounds, we can estimate $\gamma_\delta^{\rm B}$ and $\gamma_\eta^{\rm E}$.
For example, if $\delta = 10^{-6}$ and $\eta = 0.1$, we have $\gamma_\delta^{\rm B} = 0.8$ dB and $\gamma_\eta^{\rm E} = -4.8$ dB.

\begin{figure}[tb]
\begin{centering}
\psfrag{K}[Bl]{{\footnotesize{K=}}}
\includegraphics[keepaspectratio, height=70mm]{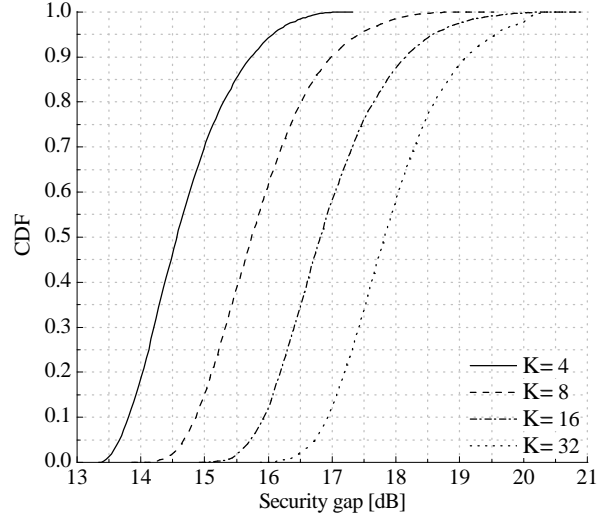}
\caption{Distribution of the security gap with $K = 4, 8, 16, 32$ parallel channels. Bob's channels are known and Alice adopts optimal power allocation.}
\label{fig:SgPowerAllocation}
\par\end{centering}
\end{figure}

If we consider that Bob's channels are known, we can use the approach in Section \ref{subsec:SgPowerAlloc}
and assume that Alice chooses the optimal power allocation strategy as given by \eqref{Pkcond}, checking that the power constraint \eqref{ac1} is verified.
Considering CPS, using \eqref{PMgee} and imposing $\myzeta = 10^{-2}$, we find the maximum value of $\alpha_{\rm E}$, as defined in Section~\ref{sec:two}, from which $\bar{\gamma}_{\max}^{\rm E}$ is obtained,
according to \eqref{ae1}. We have simulated \myhlFC{$10000$} realizations of Bob's channels, with $K = 4, 8, 16, 32$, and a maximum average power transmitted by Alice equal to $P_{\max} = \gamma_{\delta}^{\rm B}/\alpha_{\rm B}$.
The resulting CDF of the security gap is shown in Fig.~\ref{fig:SgPowerAllocation}.
The average security gap, in these four cases, is $14.68$ dB, $15.84$ dB, $16.94$ dB and $17.93$ dB for $K = 4, 8, 16$ and $32$, respectively. 

\begin{table*}%[th!]
\renewcommand{\arraystretch}{1}
\caption{Security gap $S_\myzeta$ for a ($128, 64$) eBCH coded transmission with CPS over $K$ parallel channels, with outage probability $\myzeta = 10^{-2}$. Case of equal Bob's channel gains.}
\label{tab:SgValues}
\centering
\begin{tabular}{c|c|c|c|c|c|c|c|c}
$K$ & $1$ & $2$ & $4$ & $8$ & $16$ & $32$ & $64$ & $128$ \\
\hline
$\bar{\gamma}_{\max}^{\rm E}$ & $-11.43$	dB & $-12.04$	dB & $-12.57$	dB & $-13.05$	dB & $-13.48$	dB & $-13.87$	dB & $-14.22$	dB & $-14.56$ dB \\
\hline
$S_\myzeta$ & $12.23$ dB & $12.84$ dB & $13.37$ dB & $13.85$ dB & $14.28$ dB & $14.67$ dB & $15.02$ dB & $15.36$ dB \\
\end{tabular}
\end{table*}

For the sake of comparison, we can consider an ideal scenario, in which all Bob's channel gains are equal and coincide with their mean, $H_k = \alpha_{\rm B}, k = 1,\ldots,K$.
In this case, we have $P_k = P_{\max},\ \forall k$.
This benchmark scenario is considered in Table \ref{tab:SgValues}, where we report the values of $\bar{\gamma}_{\max}^{\rm E}$ and $S_\myzeta$ for different values of $K$. We observe that, in this case, the security gap is lower than the average security gap for the case with Rayleigh distributed Bob's channels, which has been reported, for $K = 4, 8, 16$ and $32$, at the end of the previous paragraph.
%The other columns of Table \ref{tab:SgValues} will be explained in Appendix \ref{app:BobChanStat}, where also Bob's channels will be considered known only in statistical terms.

\myhl{
We compute the equivocation rate by following the derivation reported in Section \ref{subsec:EquivocationRate}. For the sake of simplicity, we consider again the case $H_k = \alpha_{\rm B}, k = 1,\ldots,K$,
yielding that uniform power allocation is the optimal solution and we have $P_kH_k = \gamma^{\rm B}$, $P_k\alpha_{\rm E} = \bar{\gamma}^{\rm E} = \gamma^{\rm B}/S_\myzeta$.
We also suppose that the minimum transmission power \eqref{Pkcond} is used to achieve the reliability target \eqref{condber1ter}.
Under these hypotheses, the constellation constrained secrecy rate becomes
%\hl{Qui $\xi$ dovrebbe essere $\epsilon$. Perch\'e c'\'e $S_\myzeta$???}
\begin{equation}
%C\sub s^{(\xi)} =  \max_{ \gamma^{\rm B} \,\leq\, \gamma^{\rm B}\sub{max}} \left[C( \gamma^{\rm B})-C\left(-\gamma\frac{\ln\xi}{KS_\myzeta}\right)\right].
C\sub s^{(\myzeta)} =  C(\gamma_\delta^{\rm B}) - C\left(- \bar{\gamma}^{\rm E} \frac{\ln\myzeta}{K}\right)\,,
\end{equation}
while the equivocation rate considering a code with rate $R_{\rm c}$ is 
\begin{equation}
R\sub e^{(\myzeta)} =  R\sub c - C\left(-\bar{\gamma}^{\rm E}\ln\left(1-(1-\myzeta)^{1/K}\right)\right), 
\end{equation}
where $C(\gamma)$ is given by
\begin{equation}
C(\gamma) = 1 - \frac{1}{\sqrt{2 \pi}} \int_{-\infty}^{+\infty} e^{-\frac{(y-\sqrt\gamma)^2}{2}} \log(1 + e^{-2y\sqrt\gamma}) {\rm d}y.
\label{eq:C}
\end{equation}
}

\begin{figure}[tb]
\begin{centering}
\includegraphics[keepaspectratio, height=65mm]{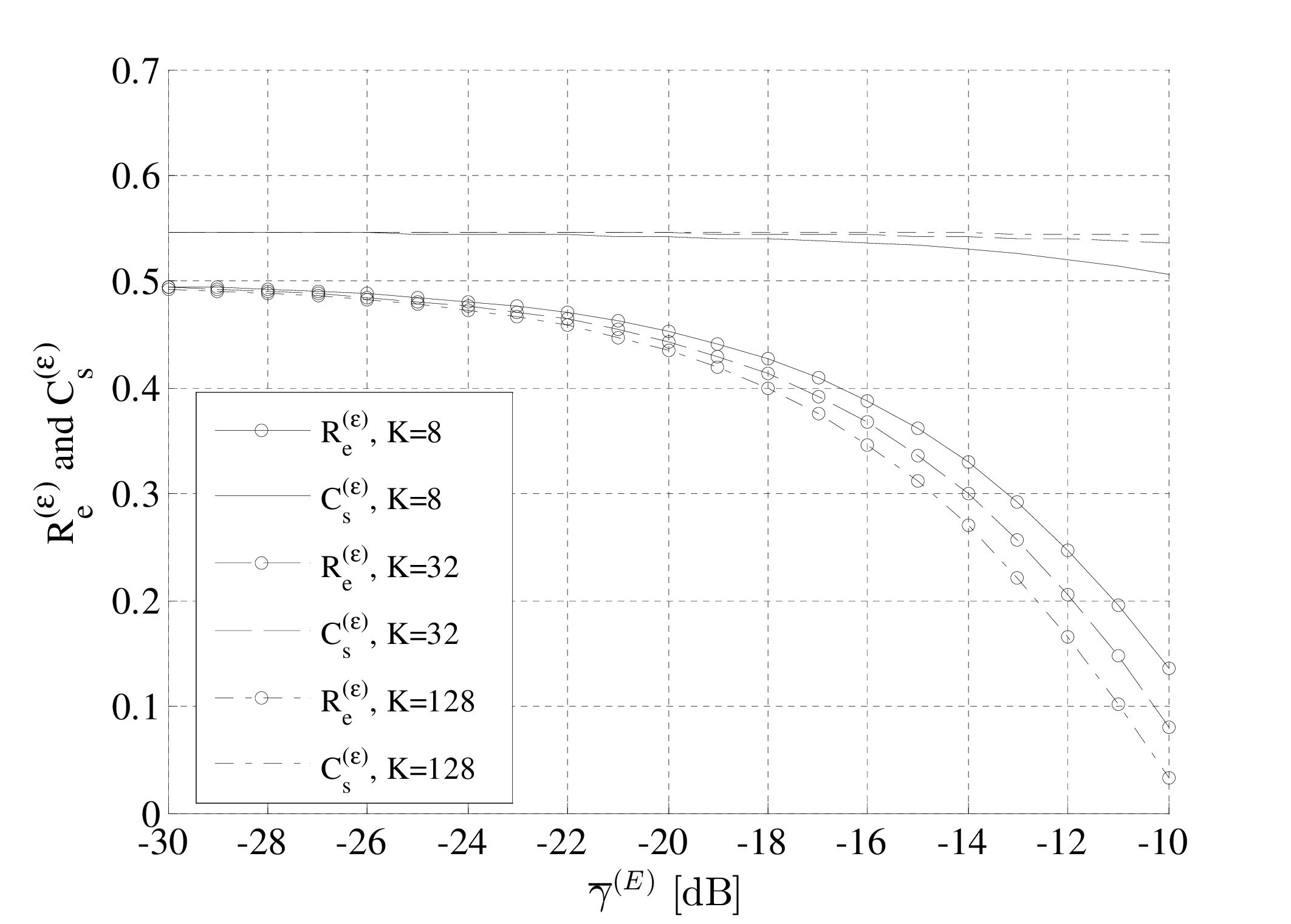}
\caption{$\myzeta$-outage equivocation rate $R\sub e^{(\myzeta)}$ and BPSK-constrained secrecy rate $C\sub s^{(\myzeta)}$ for $\gamma^{\rm B} = \gamma_\delta^{\rm B} = 0.8$ dB, $R_{\rm c} = 1/2$ and $\myzeta = 0.01$.}
\label{fig:ReCs}
\par\end{centering}
\end{figure}

By using these expressions, we have computed $R\sub e^{(\myzeta)}$ and $C\sub s^{(\myzeta)}$, as functions of $\bar{\gamma}^{\rm E}$, for $\gamma^{\rm B} = \gamma_\delta^{\rm B} = 0.8$ dB, $\myzeta = 0.01$, $R_{\rm c} = 1/2$ and some  values of $K$. Results are reported in Fig.~\ref{fig:ReCs}.
As expected, we observe that, for decreasing values of $\bar{\gamma}^{\rm E}$, the $\myzeta$-outage equivocation rate approaches the BPSK-constrained secrecy rate, and the dependence on $K$ vanishes.

\subsubsection{Secrecy rate with $K=2$ parallel channels}

In this example, we consider a simple case in which the secret message is transmitted over two channels only (i.e., $K = 2$) and the secrecy outage probability is constrained below the threshold $\epsilon = 0.01$.

%The optimal values of the transmission secrecy rates for each of the two sub-messages are determined by Theorem~\ref{th:RateAllocation} \marker{questo non mi torna perch\'e il teorema da' l'allocazione ottima delle potenze a partire da un'allocazione delle probabilit\`a di outage $\bar p_k$ dei singoli sottocanali, mentre qui l'allocazione di potenza e` fissata}, and for three different power allocation strategies: a) equal power distribution among sub-messages, b) water-filling~\cite{Cover} power distribution with respect to Bob's channels, and c) optimal power allocation, as from (\ref{potott}) and (\ref{problem2}).

We report the secrecy rates that are obtained by the optimal solution as described in (\ref{problem2}) as well as the secrecy rates achieved by two suboptimal methods. These are obtained by fixing the power allocation, in one case to an equal power distribution between the two channels, in the other by waterfilling on the legitimate receiver channels. The secrecy rates are then optimized under the given power allocation and the constraint on the secrecy outage probability.

\renewcommand\plotone[3][]{\baseplot{#1}{CPS}{contourCPS_#2_#3}}
\renewcommand\lablinea[2]{\node[anchor=south,rotate=-45] at (#1,#1) {$R\sub s = #2$};}
\begin{figure}
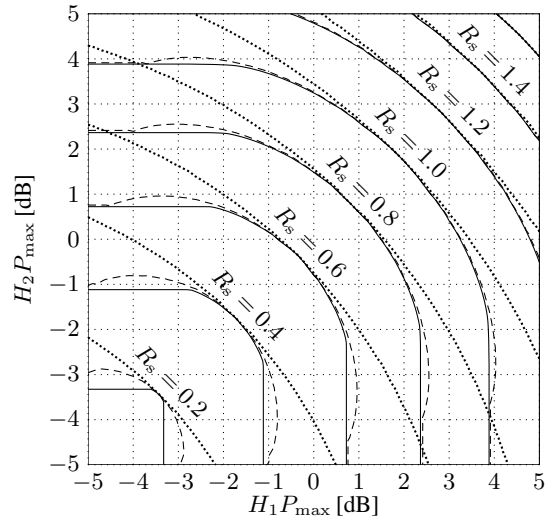

\begin{center}
\begin{footnotesize}
\Assi[f;6mm,6mm](-5,-5)(5,5){
\mygrid[dotted](-5,-5)(5,5)(1,1)
\ylabs[-5]{-5,-4,-3,-2,-1,0,1,2,3,4,5}
\xlabs[-5]{-5,-4,-3,-2,-1,0,1,2,3,4,5}
\wt(0,-5.5;t){$H_1 P\sub{max}$\,[dB]}
\wtv(-6,0;r){$H_2 P\sub{max}$\,[dB]}
\path[clip] (-5,-5) rectangle (5,5);
\foreach \lev in {02,04,06,08,10,12,14,16}
{\plotone[thick, densely dotted]{up}{\lev}
\plotone[densely dashed]{wf}{\lev}
\plotone[solid]{opt}{\lev}}
\lablinea{-3.4389}{0.2}
\lablinea{-1.7403}{0.4}
\lablinea{-0.3582}{0.6}
\lablinea{0.8028}{0.8}
\lablinea{1.8674}{1.0}
\lablinea{2.8178}{1.2}
\lablinea{3.7069}{1.4}
%\lablinea{4.5426}{1.6}
}

\end{footnotesize}
\end{center}
\caption{Contour plot of the achievable secrecy rates with with CPS under optimal (solid), waterfilling (dashed), and  equal (dotted) power allocation, for different values of Bob's channel gains. In all plots, $\epsilon = 0.01$, and $\alpha_{\rm E} P_{\rm max}=0.05$.}
\label{fig:RateContCPS}
\end{figure}

\begin{figure}
\begin{centering}
%\subfigure[]{\includegraphics[height = 0.35\textwidth, width=0.20\textwidth]{ratePowerFixedH2}}
%\subfigure[]{\includegraphics[height = 0.35\textwidth, width=0.20\textwidth]{percH2}}
\includegraphics[width=0.5\textwidth]{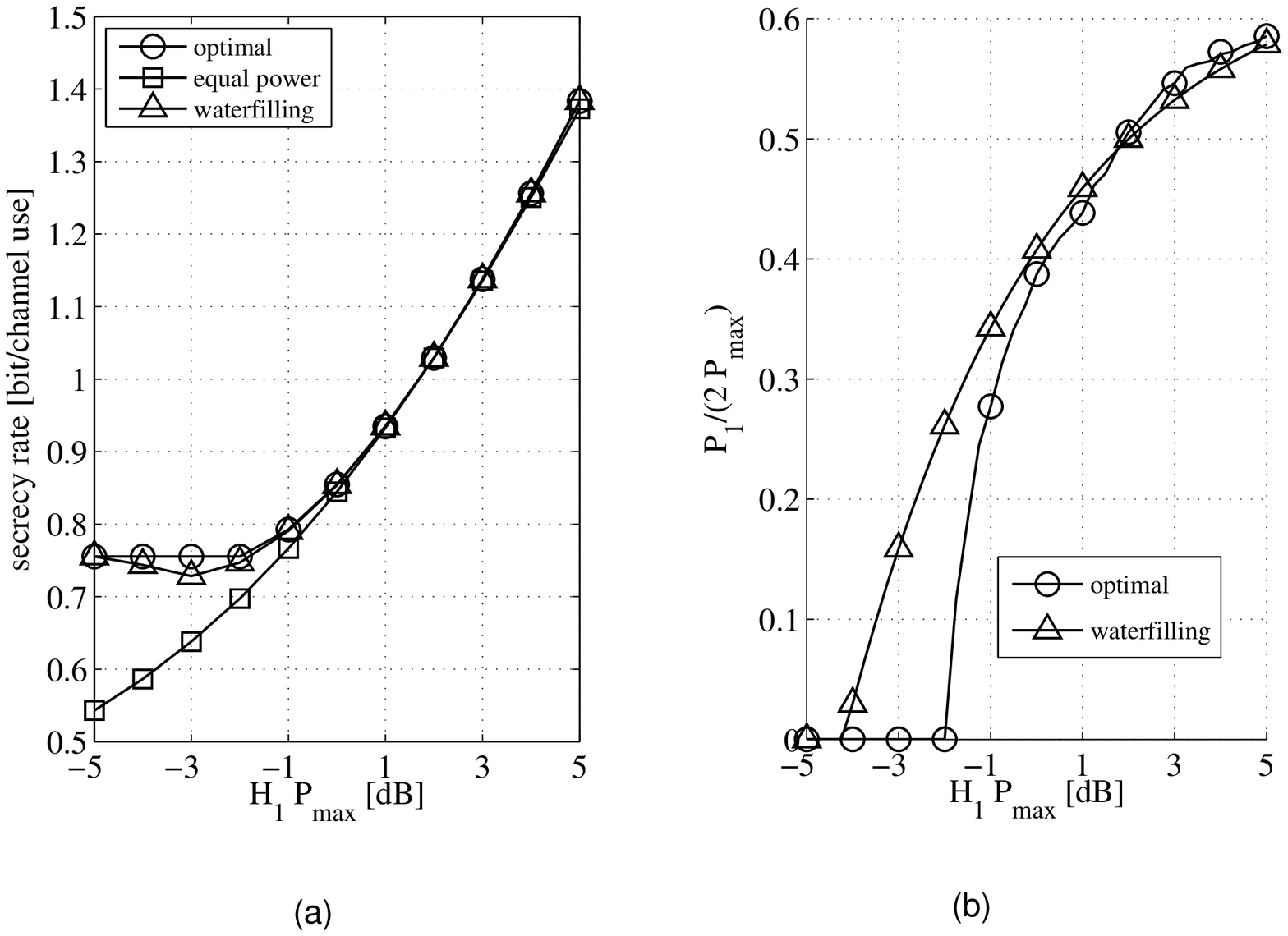}
\caption{(a) Achievable secrecy rates with CPS and (b) fraction $P_1/(2P\sub{max})$ of the available power that is allocated to $k = 1$. In all plots, $\epsilon = 0.01$, $\alpha_{\rm E} P_{\rm max}=0.05$ and $H_2P_{\max} = 2$ dB.}
\label{fig:RateCont2}
\par\end{centering}
\end{figure}

Fig.~\ref{fig:RateContCPS} shows the contour lines of the secrecy rates obtained with the different power allocations described above, as a function of the power gains of Bob's channels. The eavesdropper average power gain is such that $\alpha_{\rm E} P_{\rm max}=0.05$ for each channel. As expected from the symmetry of the problem, the three strategies provide similar performance when $H_1$ and $H_2$ are close to each other, as all three methods equally divide the power between the two channels. On the other hand, when $H_1$ and $H_2$ are very different, the optimal solution is to load all power on the stronger channel while  waterfilling is suboptimal, and equal power allocation achieves a much lower rate. Waterfilling loses against the optimal solution in the intermediate region, as it is possible to observe from Fig.~\ref{fig:RateCont2}(a), in which the secrecy rates are shown for a specific value of Bob's gain in the second channel, i.e., $H_2 P_{\rm max}=2$\,dB. The loss can also be seen (although it is not shown here) to be increasing with the values of $\alpha_{\rm E}$, since as $\alpha_{\rm E}$ decreases the constraint on secrecy becomes less stringent than that on reliability, and waterfilling becomes more effective.
This effect is explained by observing that waterfilling allocates power to a channel when its gain is above a certain threshold to guarantee a benefit in terms of transmission rate without secrecy constraints. However, when a constraint is imposed on the secrecy outage probability, this threshold increases.
This is also observed in Fig.~\ref{fig:RateCont2}(b), in which the fraction of power allocated to the first channel by the two non-uniform methods is reported. \myhlN{We see that, when $H_1$ is small, the optimal power allocation provides the first channel with a lower fraction of power compared to waterfilling. In particular, in order to allocate power to the first channel, the optimal joint rate/power allocation method requires a significantly higher average received power than that required by the waterfilling solution.}

\subsection{Coding Across Sub-messages}

\subsubsection{Security gap with eBCH coding}

\myhl{
Let us consider the case of $K=128$ parallel channels and the same code used in Section \ref{ssCPS}, having $N = 128$ and rate $R_{\rm c} = 1/2$.
The value of $\bar{\gamma}_{\max}^{\rm E}$ can be estimated through \eqref{ae2}, which provides the same result
already computed for this code with CPS over $K=128$ parallel channels, due to the use of the lower bound \eqref{lbPE}.
Therefore, by considering $\myzeta = 10^{-2}$, we obtain $S_\myzeta = 15.36$ dB, as reported in Table \ref{tab:SgValues}.

When Bob's channel is also known only in statistical terms, we can use the approach described in Appendix \ref{app:BobChanStat}
to estimate the security gap.
This way, and by using the upper bound \eqref{ubPB} for estimating $\bar{\gamma}_{\min}^{\rm B}$, we obtain that the
system requires a security gap equal to $56.41$ dB, which is the same as with CPS and $K=128$.
Nevertheless, we can avoid to use the bound \eqref{ubPB} by exploiting, for the case of CAS, the per-realization method
described in {\cite{Snow2006}}.
This way, as detailed in Appendix \ref{app:BobChanStat}, we obtain a security gap equal to $18.21$ dB, which highlights the
superiority of CAS over CPS in these conditions.
}

\subsubsection{Secrecy rate with $K=2$ parallel channels}

\renewcommand\plotone[3][]{\baseplot{#1}{CAS}{contourCAS_#2_#3}}
\renewcommand\lablinea[2]{\node[anchor=south,rotate=-45] at (#1,#1) {$R\sub s = #2$};}
\begin{figure}
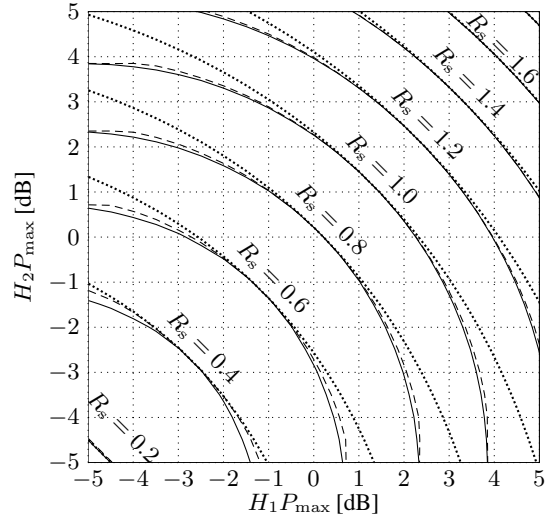

\begin{center}
\begin{footnotesize}
\Assi[f;6mm,6mm](-5,-5)(5,5){
\mygrid[dotted](-5,-5)(5,5)(1,1)
\ylabs[-5]{-5,-4,-3,-2,-1,0,1,2,3,4,5}
\xlabs[-5]{-5,-4,-3,-2,-1,0,1,2,3,4,5}
\wt(0,-5.5;t){$H_1 P\sub{max}$\,[dB]}
\wtv(-6,0;r){$H_2 P\sub{max}$\,[dB]}
\path[clip] (-5,-5) rectangle (5,5);
\foreach \lev in {02,04,06,08,10,12,14,16,18}
{\plotone[thick, densely dotted]{up}{\lev}
\plotone[densely dashed]{wf}{\lev}
\plotone[solid]{opt}{\lev}}
\lablinea{-4.504}{0.2}
\lablinea{-2.7275}{0.4}
\lablinea{-1.1794}{0.6}
\lablinea{0.1077}{0.8}
\lablinea{1.2262}{1.0}
\lablinea{2.2335}{1.2}
\lablinea{3.1610}{1.4}
\lablinea{3.9}{1.6}
%\lablinea{4.8469}{1.8}
}

\end{footnotesize}
\end{center}
\caption{Contour plot of the achievable secrecy rates with with CAS under optimal (solid), waterfilling (dashed), and  equal (dotted) power allocation, for different values of Bob's channel gains. In all plots, $\epsilon = 0.01$, and $\alpha_{\rm E} P_{\rm max}=0.05$.}
\label{fig:RateContCAS}
\end{figure}

\begin{figure}
\begin{centering}
%\subfigure[]{\includegraphics[height = 0.35\textwidth, width=0.20\textwidth]{Fig7a_addendum}}
%\subfigure[]{\includegraphics[height = 0.35\textwidth, width=0.20\textwidth]{Fig7b_addendum}}
\includegraphics[width=0.5\textwidth]{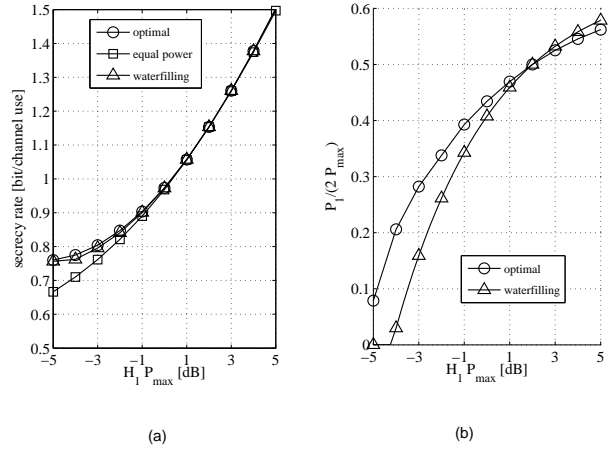}
\caption{(a) Achievable secrecy rates with CAS and (b) fraction $P_1/(2P\sub{max})$ of the available power that is allocated to $k = 1$. In all plots, $\epsilon = 0.01$, $\alpha_{\rm E} P_{\rm max}=0.05$ and $H_2P_{\max} = 2$ dB.}
\label{fig:RateContCA2}
\par\end{centering}
\end{figure}

In this example, we assess the secrecy rate of the CAS scheme and compare it with the corresponding secrecy rate achieved by CPS for the simple case of $K=2$ parallel channels.
Fig.~\ref{fig:RateContCAS} shows the contour lines of the secrecy rate (\ref{eq:rateCA}) obtained within the same two-channel scenario and for the three power allocation strategies considered in Section~\ref{ssCPS}. 
\myhlF{
Due to its increased flexibility, we expect that CAS outperforms CPS when achievable rates are considered. Indeed, this is confirmed by comparing Fig. \ref{fig:RateContCAS} with Fig. \ref{fig:RateContCPS}. Still, note that their \myhlFC{performances} in the considered simulation scenario are quite close, so that other implementation issues may guide the choice between the two schemes. For example, CAS is more robust against imperfect power allocation, as we note that the loss incurred by equal power allocation and waterfilling with respect to the optimal solution is almost negligible for a wide range of channel gains (as can be also observed in Fig.~\ref{fig:RateContCAS}). However, other issues may be relevant for a complete comparison, as those mentioned at the end of Section \ref{sec:two}. On the other hand, Fig.~\ref{fig:RateContCA2}(b) shows that, opposite to CPS, when $H_1$ is small, the optimal power allocation for CAS provides the first channel with a higher fraction of power compared to waterfilling.
}
%Perhaps surprisingly, and in contrast with the results obtained for different parallel channels scenarios in the literature {\cite{Khisti08, Gopala08, Liang08}}, by comparing Fig. \ref{fig:RateContCAS} with Fig. \ref{fig:RateContCPS} we observe that CAS yields slightly higher secrecy rates than CPS. Moreover, CAS is more robust against imperfect power allocation, as we note that the loss incurred by equal power allocation and waterfilling with respect to the optimal solution is almost negligible for a wide range of channel gains (This can be also observed in Fig.~\ref{fig:RateContCAS}). It must be noted that such a favorable behavior of CAS over CPS does not permit us to conclude that CAS is better than CPS. The conclusion, in fact, has been derived for a specific scenario and, even more important, other issues must be considered in the comparison, like those mentioned at the end of Section \ref{sec:two}. On the other hand, Fig.~\ref{fig:RateContCA2}(b) shows that, opposite to CPS, when $H_1$ is small, the optimal power allocation for CAS provides the first channel with a higher fraction of power compared to waterfilling.

\subsubsection{Channel selection and uniform power allocation}

\begin{figure}
\begin{centering}
\subfigure[]{
\includegraphics[keepaspectratio, width=0.48\textwidth]{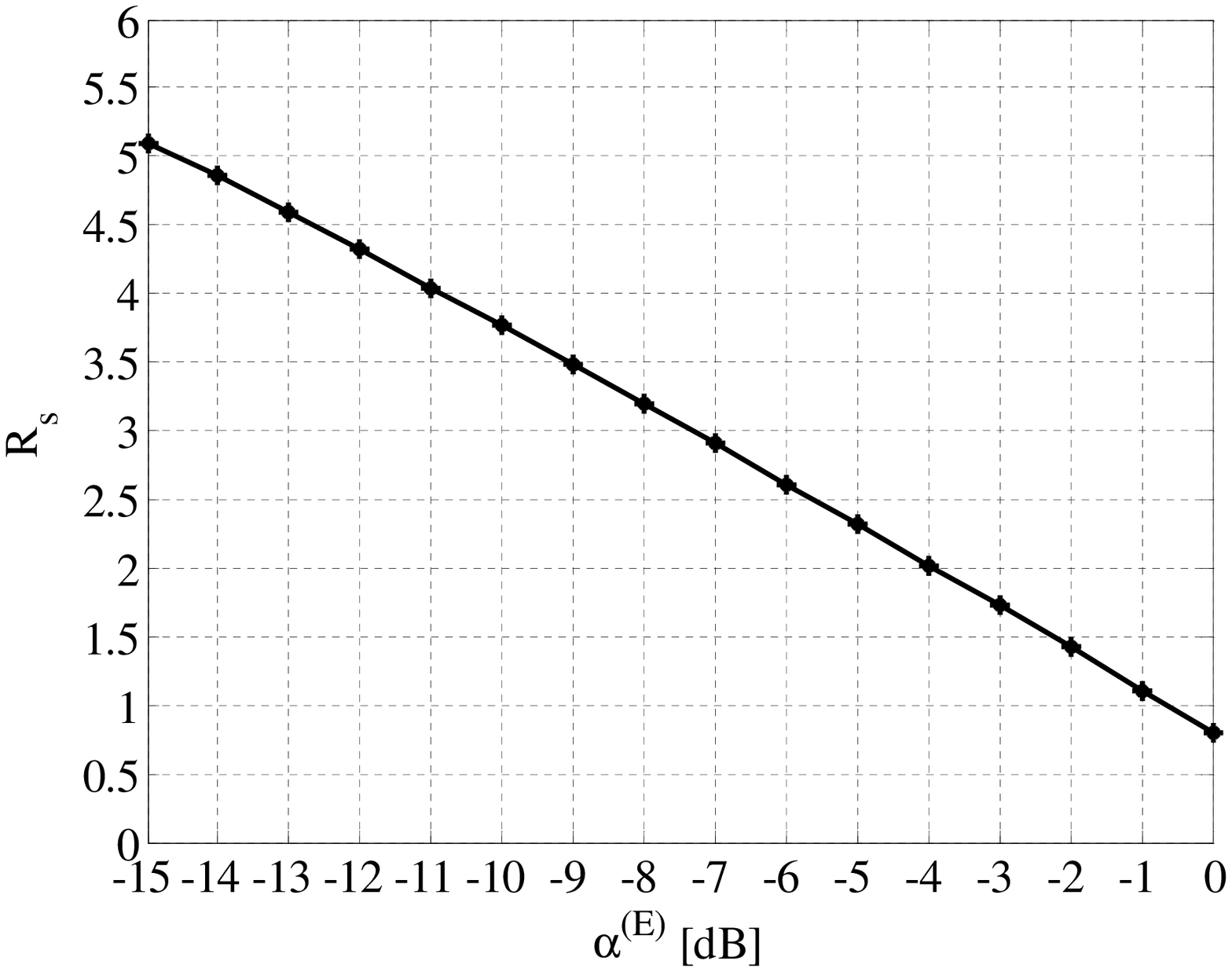}}
\subfigure[]{\includegraphics[keepaspectratio, width=0.48\textwidth]{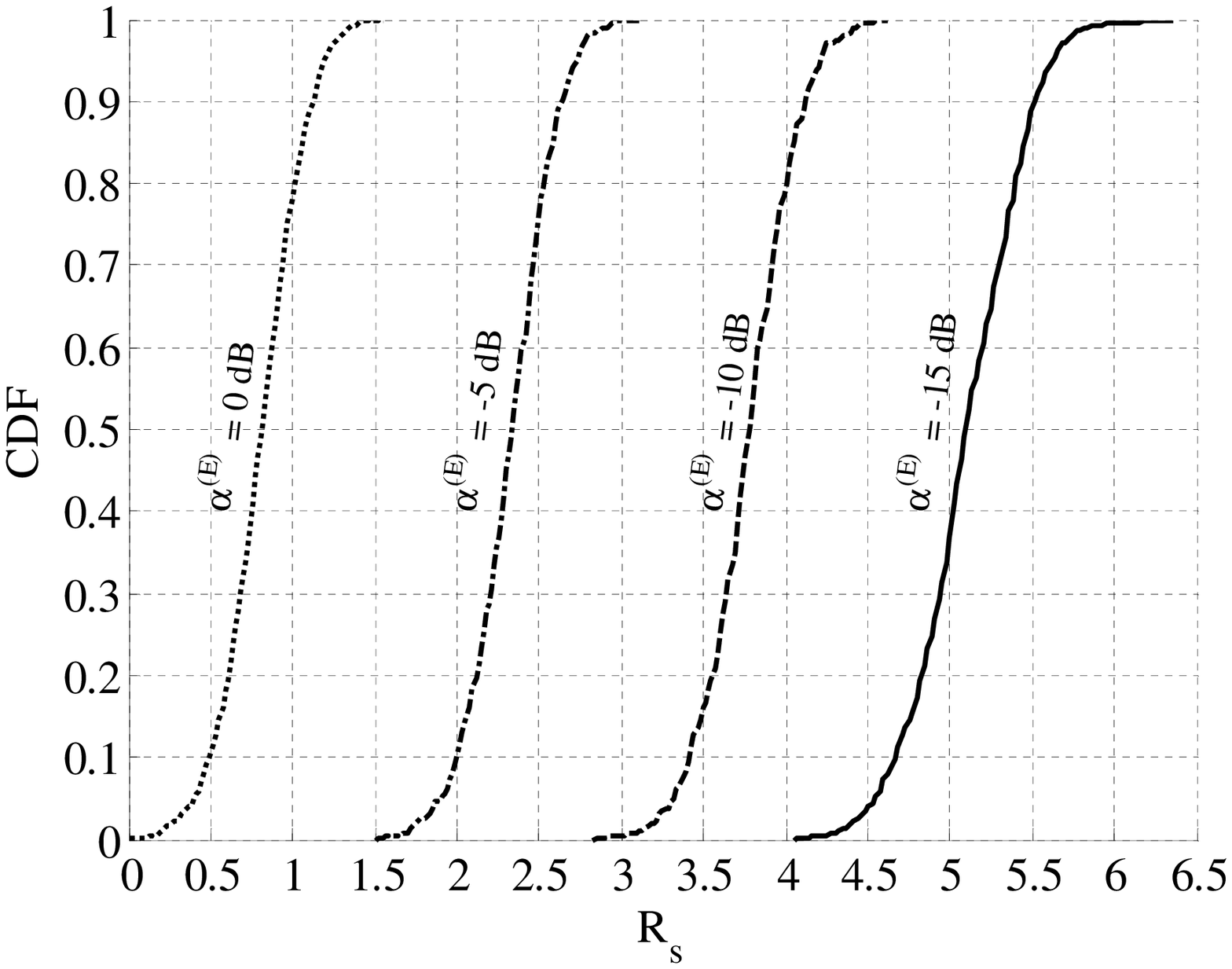}}
\caption{Outage secrecy rates with selection of $K' \le K = 48$ sub-messages and uniform power allocation among them: (a) mean $\epsilon$-outage secrecy rate and (b) \ac{CDF} of secrecy rates due to the statistics of Bob's channel gains around $\alpha_{\rm B}$.}
\label{fig:scge}
\end{centering}
\end{figure}

As a more practical example of the use of CAS, we have also considered the case of $K=48$ channels. Since the computation of the optimal power allocation in this case is infeasible, we have considered a suboptimal approach, in which only a subset of $K' \le K$ channels (those with the highest $H_k$) are used and the available power is allocated uniformly among them. 
By imposing an outage probability $\epsilon = 0.01$, and considering $P\sub{max} = 3.8$ dB and $\alpha_{\rm B} = 1$, Fig.~\ref{fig:scge}(a) shows the maximum (over all values of $K' \le K$) mean outage secrecy rate $R\sub s = \frac1K\sum_k R_k$ as a function of $\alpha_{\rm E}$. As expected, from the figure we observe that, as $\alpha_{\rm E}$ increases, the mean outage secrecy rate decreases. This behavior is confirmed by Fig.~\ref{fig:scge}(b), that shows \ac{CDF}s of the outage secrecy rates due to the statistics of Bob's channel, for different values of $\alpha_{\rm E}$.  

\section{Conclusions}
\label{sec:six}

In this paper we have characterized the performance of secret transmissions over parallel channels, under the assumption of
knowing the Alice-Bob channel and having only a statistical description of the Alice-Eve channel.
We have used a set of metrics that allow studying the problem both from the theoretical standpoint and by considering practical coded transmission schemes.
We have derived bounds on the achievable outage secrecy rates (using ideal codes), and studied the effect of power allocation on the
secrecy performance.
The definitions of security gap and equivocation rate have been extended to
this scenario, and we have used them to assess the requirements for achieving
security when practical codes are adopted.

\begin{comment}
\section*{Acknowledgment}
The authors are very grateful to the anonymous reviewers for their valuable comments and suggestions, that have contributed to improve significantly the contents and presentation of the paper.
\end{comment}

\appendices

\section{Proof of Theorem~\ref{th:RateAllocation}}
\label{par:ProofRate}

From (\ref{poutset}) we have
\begin{equation}
  \frac{1+H_k P_k}{P_k 2^{R_k}} - \frac{1}{P_k}  = - \alpha_{\rm E}\ln \bar{p}_k
\end{equation}
%\begin{equation}
%   2^{R_k} =  \left[\left(- \alpha_{\rm E}\ln \bar{p}_k + \frac{1}{P_k} \right)  \frac{P_k}{1+H_k P_k}\right]^{-1}
%\end{equation}
which can be rewritten as 
\begin{equation}
   2^{R_k} =  \frac{1+H_k P_k}{\bar{u}_k P_k + 1  }\,.
   \label{2rk}
\end{equation}
From (\ref{2rk}) we immediately obtain the second result of the theorem.

By the Karush-Kuhn-Tucker (KKT) conditions, problem (\ref{problem}) subject to power constraint (\ref{ac1}) can be written as 
\begin{equation}
\max_{\bm{P}, \nu} \sum_{k=1}^K \left[\log \frac{1+H_k P_k}{\bar{u}_k P_k + 1  } - \nu (P_k - P_{\rm max})\right]\,.
\end{equation}
Setting to zero the derivative with respect to $P_k$ we obtain
\begin{equation}
 \frac{\bar{u}_k P_k + 1  }{1+H_k P_k} \left[  \frac{H_k }{\bar{u}_k P_k + 1  } -  \frac{\bar{u}_k(1+H_k P_k)}{(\bar{u}_k P_k + 1 )^2}\right] - \nu = 0
\end{equation}
which can be rewritten as 
\begin{equation}
\bar{u}_k\nu  H_k P_k^2 + \nu  (\bar{u}_k + H_k) P_k +\nu  -   H_k +  \bar{u}_k = 0.
\label{eq18}
\end{equation}
Now, from (\ref{eq18}), if $\nu  -   H_k +  \bar{u}_k <0$ we obtain (\ref{potott}).

\section{Proof of (\ref{psfinal})}
\label{prooffinal}

From (\ref{eq:psCond}) and (\ref{defPhi}) we can rewrite $\myhlF{p_s(\bm{P}, \erreuno;\bm{H})}$ as 
\begin{equation}
\begin{split}
\myhlF{p_s(\bm{P}, \erreuno;\bm{H})} = & \P{\prod_{k=1}^K (1 + G_k P_k) \geq \myhlFC{\Phi(\bm{P}, \erreuno;\bm{H})}}\,. \\
\end{split}
\end{equation}
Defining $\beta = \prod_{k=1}^K (1 + G_k P_k)$ we have
\begin{equation}
\myhlF{p_s(\bm{P}, \erreuno;\bm{H})} =  1 -\int_{1}^{\myhlFC{\Phi(\bm{P}, \erreuno;\bm{H})}} p_\beta(a) da\,
\label{eqs}
\end{equation}
with $p_\beta(a)$ the PDF of $\beta$, which has been computed in {\cite{Yilmaz}}. We recall the definition of the generalized Fox H-function {\cite{Yilmaz}} 
\begin{equation}
\begin{split}
\mathcal H_{p,q}^{m,n}&\left[r\left|\begin{array}{c} \{a_i, c_i, A_i\} \\ \{b_j, d_j, B_j\} \end{array} \right.\right] = \\
& \frac{1}{2\pi {\rm i}} \oint_{\mathcal C} M_{p,q}^{m,n}\left[s\left|\begin{array}{c} \{a_i, c_i, A_i\} \\ \{b_j, d_j, B_j\} \end{array} \right.\right] r^{-s} {\rm d}s\,,
 \end{split}
\label{defH}
\end{equation} 
where $\mathcal C$ is a contour in the complex plane from $\omega - {\rm i} \infty$ to $\omega + {\rm i} \infty$ (where i is the imaginary unit) such that $(b_i + k)/d_i$ and $(a_i -1-k)/c_i$ (with $k$ non-negative integer) lie to the right and left of $\mathcal C$, respectively, and
\begin{equation}
\begin{split}
M_{p,q}^{m,n}&\left[s\left|\begin{array}{c} \{a_i, c_i, A_i\} \\ \{b_j, d_j, B_j\} \end{array} \right.\right] = \\
&\frac{  \prod_{j=1}^m \hat{\Gamma}(b_j+d_j s, B_j) \prod_{i=1}^n \hat{\Gamma}(1-a_i-c_is, A_i)} {  \prod_{i=n+1}^p \hat{\Gamma}(a_i+c_i s, A_i) \prod_{j=m+1}^q \hat{\Gamma}(1-b_j-d_j s, B_j)}
\end{split}
\label{Mdef}
\end{equation}
is the Mellin transform of the generalized Fox H-function, where $\hat{\Gamma}(\cdot, \cdot)$ is the upper incomplete Gamma function
\begin{equation}
\hat{\Gamma}(s, a) = \int_a^\infty t^{s-1} e^{-t} {\rm d} t\,,
\end{equation}
and an empty product is taken to be one. We have {\cite{Yilmaz}}
\begin{equation*}
p_\beta(a) = \frac{\myzeta(\bm{P})}{\phi(\bm{P})} \mathcal H_{0, K}^{K,0} \left[ \left. \frac{a}{\phi(\bm{P})}\right|\begin{array}{c}  \{-, -, -\} \\ \{0, 1, (P_k \alpha_{\rm E})^{-1}\}_{k = 1, \ldots, K} \end{array} \right]\,,
\end{equation*}
for $a \geq 1$ and $p_\beta(a) = 0$ otherwise. The notation $\{-, -, -\}$ means that the coefficients are absent. Now, by observing that 
\begin{equation}
\int_{1}^q t^{-s} {\rm d}t= \frac{ q^{1-s} -1}{1-s} = (q^{1-s} - 1) \frac{\hat{\Gamma}(1-s, 0)}{\hat{\Gamma}(2-s, 0)}
\label{intsmart}
\end{equation}
and inserting the integral of (\ref{eqs}) into (\ref{defH}) and using (\ref{intsmart}) together with (\ref{Mdef}) we obtain (\ref{psfinal}).

\section{Security gap for Bob's channel known only in statistical terms}
\label{app:BobChanStat}

When Bob's channels are known only in statistical terms, we consider an outage approach also for Bob in order to define the security gap.
In fact, $p^{\rm B}$ is a random variable, whose distribution is uniquely determined by the average SNR.
The average \ac{SNR} of the Alice-Bob channel is
\begin{equation}
\bar{\gamma}^{\rm B} = \frac{1}{K} \sum_k \alpha_{\rm B} P_k,
\label{eq:GammaBarB}
\end{equation}
under the assumption of \ac{i.i.d.} Rayleigh channels.

We are interested in finding the minimum value of $\bar{\gamma}^{\rm B}$, denoted as $\bar{\gamma}_{\min}^{\rm B}$,
for which the probability that \eqref{condber1ter} does not hold is not greater than $\omega$, i.e., for CPS
\begin{subequations}
\label{ab1}
\begin{equation}
\bar{\gamma}_{\min}^{\rm B} = 
\frac{1}{K} \sum_k P_k  \min \{\alpha_{\rm B}: \P{\cup_{k=1}^K\{p_k^{\rm B} > \delta\}} \le \omega\}
\end{equation}
and for CAS
\begin{equation}
\bar{\gamma}_{\min}^{\rm B} = 
\frac{1}{K} \sum_k P_k  \min \{\alpha_{\rm B}: \P{p^{\rm B} > \delta} \le \omega\}\,,
\end{equation}
\end{subequations}
where $p_k^{\rm B}$ is Bob's \ac{CER} on each channel.
Then the $(\omega,\myzeta)$ security gap in this case is defined as 
\begin{equation}
S_{\omega,\myzeta} = \frac{\bar{\gamma}_{\min}^{\rm B}}{\bar{\gamma}_{\max}^{\rm E}}\,,
\label{Sgomegazeta}
\end{equation}
where $\bar{\gamma}_{\max}^{\rm E}$ is given by \eqref{ae1}.
The value of $\bar{\gamma}_{\min}^{\rm B}$ is computed next for CPS and CAS.

\paragraph*{Computation of $\bar{\gamma}_{\min}^{\rm B}$ for CPS} When CPS is considered, we must impose that $p_k^{\rm B} \le \delta$. Therefore, Bob's outage probability is
\begin{equation}
\begin{split}
\P{\cup_{k=1}^K\{p_k^{\rm B} > \delta\}} = 1 - \prod_{k=1}^K \left(1 - \P{p_k^{\rm B} > \delta}\right) = \\
1- \prod_{k=1}^K (1 - \P{P_kH_k < \gamma_\delta^{\rm B}(k)})\,.
\end{split}
\label{Pbout}
\end{equation}
From the Rayleigh distribution assumption, we have $\P{P_kH_k < \gamma_\delta^{\rm B}(k)} = 1 - \exp \left( - \frac{\gamma_\delta^{\rm B}(k)}{P_k \alpha_{\rm B}} \right)$, and by equating the r.h.s. of \eqref{Pbout} to $\omega$, we can derive the value of $\alpha_{\rm B}$ and the corresponding $\bar{\gamma}_{\min}^{\rm B}$.

By using the same assumptions as in Section \ref{sec:five}, i.e., by considering the use of BPSK and 
of the same eBCH code with length $N = 128$ and rate $R_{\rm c} = 1/2$ on all channels,
we have computed $\bar{\gamma}_{\min}^{\rm B}$ for the case of CPS.
The results, and the corresponding values of $S_{\omega,\myzeta}$ obtained by using the values of $\bar{\gamma}_{\max}^{\rm E}$ given in Table \ref{tab:SgValues}, 
are reported in Table \ref{tab:SgValuesBis}.
As expected, when Bob's channels are known only in statistical terms, the values of the security gap needed to ensure conditions \eqref{condber} are significantly higher than those for the case in which Bob's channels are known exactly, which have been reported in Table \ref{tab:SgValues}.

\begin{table*}[th!]
\renewcommand{\arraystretch}{1}
\caption{Security gap $S_{\omega,\myzeta}$ for a ($128, 64$) eBCH coded transmission with CPS over $K$ parallel channels, with outage probabilities $\omega = \myzeta = 10^{-2}$.}
\label{tab:SgValuesBis}
\centering
\begin{tabular}{c|c|c|c|c|c|c|c|c}
$K$ & $1$ & $2$ & $4$ & $8$ & $16$ & $32$ & $64$ & $128$ \\
\hline
$\bar{\gamma}_{\min}^{\rm B}$ & $20.78$	dB & $23.79$dB & $26.80$	dB & $29.81$ dB & $32.82$ dB & $35.83$ dB & $38.84$ dB &  $41.85$ dB \\
\hline
$S_{\omega,\myzeta}$ & $32.21$ dB & $35.83$ dB & $39.37$ dB & $42.86$ dB & $46.30$ dB & $49.70$ dB & $53.06$ dB & $56.41$ dB \\
\end{tabular}
\end{table*}

\paragraph*{Computation of $\bar{\gamma}_{\min}^{\rm B}$ for CAS}

\begin{figure}[tb]
\begin{centering}
\includegraphics[keepaspectratio, height=70mm]{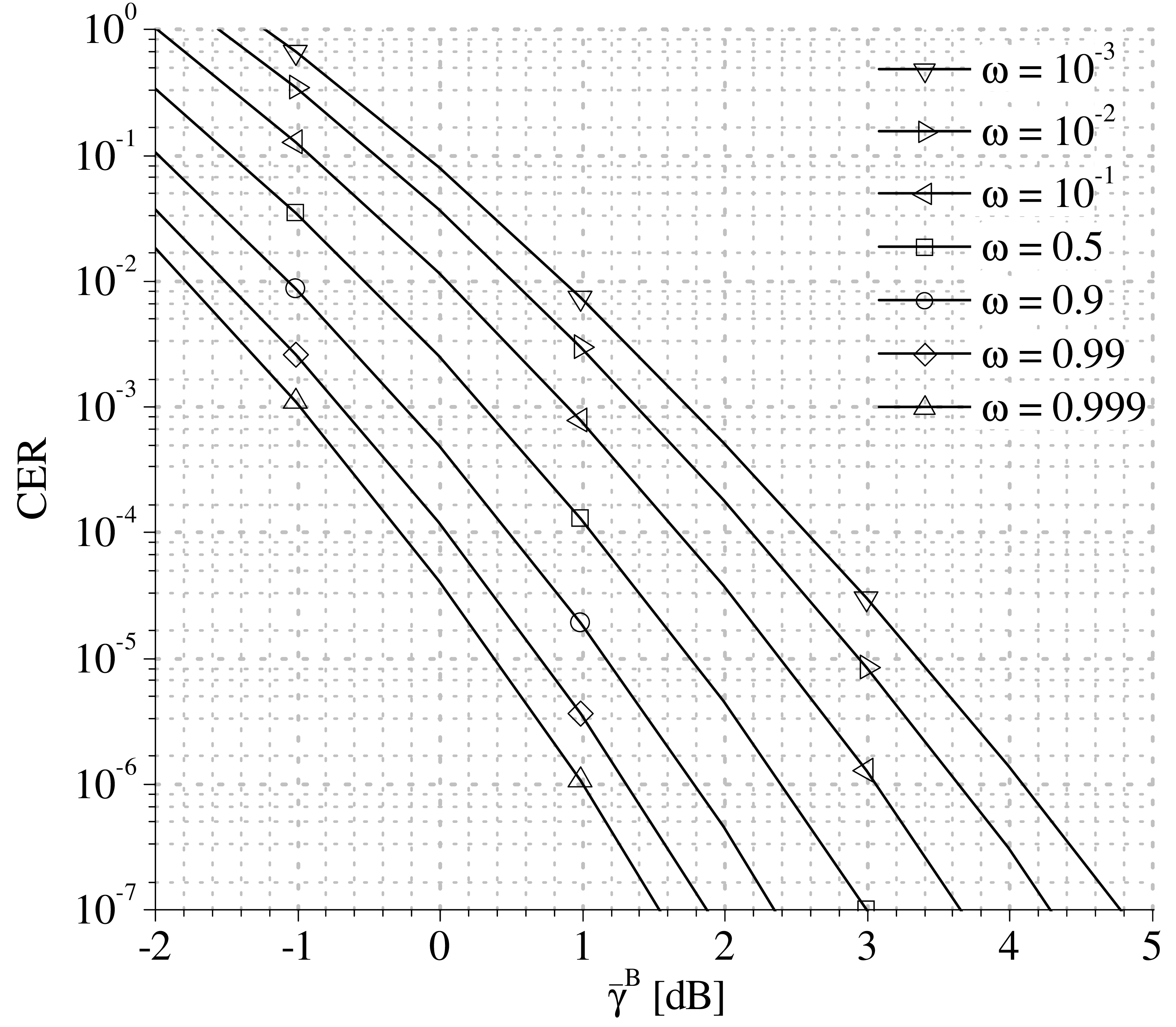}
\caption{Upper bound on the \ac{CER} estimated through the per-realization analysis for the eBCH code with length $N = 128$, rate $R_{\rm c} = 1/2$ and several values of Bob's outage $\omega$.}
\label{fig:eBCHoutage}
\par\end{centering}
\end{figure}

Similarly to what has been done for Eve in Section \ref{subsec:GammaMaxE}, for CAS we are interested in finding a worst-case estimate of Bob's error probability.
As a closed form expression is not available, we resort to an upper bound. In particular we have 
\begin{equation}
\begin{split}
\P{E^{\rm B}|P_1H_1 = \gamma_1,\ldots,P_KH_K = \gamma_K} \leq \\
\P{E^{\rm B}|P_1H_1 = \gamma_m,\ldots,P_KH_K = \gamma_m}
%p^{\rm B} = \widehat{\pi}_N^{\rm B} \le \pi_m^{\rm B},
\end{split}
\label{ubPB}
\end{equation}
where $m = \arg\min_k \gamma_k$ is the index of the channel with the minimum \ac{SNR}. 
\myhl{
Hence, a sufficient condition for \eqref{condber1ter} is that $\min_k P_kH_k \geq \max_k \gamma_\delta^{\rm B}(k)$, 
and we can replace \eqref{ab1} with 
$\bar{\gamma}_{\min}^{\rm B} = \frac{1}{K} \sum_k P_k  \min \{\alpha_{\rm B}: \P{\min_k\{P_kH_k\} < \max_k \gamma_\delta^{\rm B}(k)} \le \omega\}$.
Moreover, we have
\begin{equation}
\begin{split}
\P{\min_k\{P_kH_k\} < \max_k \gamma_\delta^{\rm B}(k)} = \\ \P{\displaystyle\bigcup_{k=1}^K P_kH_k < \max_k \gamma_\delta^{\rm B}(k)} = \\
1- \prod_{k=1}^K (1 - \P{P_kH_k < \max_k \gamma_\delta^{\rm B}(k)})\,.
\end{split}
\label{Pmlee}
\end{equation}

So, in the special case in which $\gamma_\delta^{\rm B}(1) = \gamma_\delta^{\rm B}(2) = \ldots = \gamma_\delta^{\rm B}(K) = \gamma_\delta^{\rm B}$,
we obtain again the same expression for both CPS and CAS scenarios.
However, for CPS it provides exact results, while for CAS it is due to the use of the upper bound \eqref{ubPB}.

Let us consider an example with $K=128$ parallel channels, over which the same ($128, 64$) eBCH code considered in Section \ref{sec:five}
is used to implement CAS.
In this case, we have $\bar{\gamma}_{\max}^{\rm E} = -14.56$ dB, $\bar{\gamma}_{\min}^{\rm B} = 41.85$ dB and $S_{\omega,\myzeta} = 56.41$ dB,
as it results from Tables \ref{tab:SgValues} and \ref{tab:SgValuesBis}.

In order to avoid resorting to the upper bound \eqref{ubPB} for estimating $\bar{\gamma}_{\min}^{\rm B}$, we can use the per-realization
method described in {\cite{Snow2006}}.
This method provides an estimate of the \ac{CER} achieved by a given code when each coded bit is transmitted over a channel with
a different gain, and the channel gains are Rayleigh distributed. This situation exactly models the CAS scenario we consider, and the
estimate so found is tight for \ac{ML}-like decoders and high \ac{SNR} values, that matches with Bob's condition.
Therefore, we have applied this method by computing all the $243840$ codewords with weight $22$ in the ($128, 64$) eBCH code, according to {\cite{Baumert1977}}.
The results obtained are reported in Fig. \ref{fig:eBCHoutage} in terms of the estimated \ac{CER} as a function of $\bar{\gamma}^{\rm B}$, for several
values of Bob's outage probability $\omega$.
Based on these results, we get that $p^{\rm B} \le 10^{-6}$ for $\bar{\gamma}^{\rm B} \ge \bar{\gamma}_{\min}^{\rm B} = 3.65$ dB and $\omega = 10^{-2}$ .
Therefore, a tighter estimate of the security gap in this case is $S_{\omega,\myzeta} = 18.21$ dB.
}

\section{On the computation of $\gamma_\eta^{\rm E}$}
\label{appeta}

We assume that Eve uses ML decoding, which represents the most dangerous condition for the legitimate receiver.
In order to assess Eve's error rate, we use Shannon's \ac{SPB} on the error probability of a coded transmission with ML decoding {\cite{Shannon1959}}, which is the tightest one for high error rate values, at which Eve is supposed to operate.

By Shannon's \ac{SPB} on the block error probability under ML decoding, the error probability at Eve's over a single channel with \ac{SNR} $\gamma$ is bounded by {\cite{Wiechman2008}}
\begin{equation}
p^{\rm E}(\gamma) > P_{\mathrm{SPB}}(N, \theta, A),
\label{eq:ShannonLB}
\end{equation}
where $P_{\mathrm{SPB}}(N,\theta,A)$ is the probability that the received
vector falls outside the $N$-dimensional circular cone of half angle $\theta$ whose main axis passes through both the origin and the point corresponding to the transmitted signal {\cite{Wiechman2008}}.
In \eqref{eq:ShannonLB}, $A = \sqrt{2 R_{\rm c} \gamma}$, where $R_{\rm c}$ is the code rate.

The tightest lower bound on the error probability is achieved for $\theta_1(N,R_{cn})$ such that
\begin{equation}
\frac{\Omega_N(\theta_1(N,R_{cn}))}{\Omega_N(\pi)} = \exp(-N R_{cn}),
\end{equation}
where $R_{cn}$ is the code rate in nats per channel use, $\Omega_N(\theta) = \frac{2\pi^{(N-1)/2}}{\Gamma((N-1)/2)} \int_0^\theta{(\sin \phi)^{N-2}} {\rm d}\phi$,
$\Omega_N(\pi) = \frac{2\pi^{N/2}}{\Gamma(N/2)}$, and $\Gamma(\cdot)$ denotes the Gamma function. From {\cite{Wiechman2008}} we have
\begin{equation}
\begin{split}
P&_{\mathrm{SPB}}(N,\theta,A) = \frac{(N-1)\exp\left(-\frac{NA^2}{2}\right)}{\sqrt{2\pi}} \times\\
& \int_\theta^{\frac{\pi}{2}}{(\sin \phi)^{N-2} f_N\left( \sqrt{N} A \cos \phi \right)} {\rm d}\phi + Q\left(\sqrt{N}A \right),
\end{split}
\end{equation}
where $f_N(x) = \sum_{j=0}^{N-1} \exp\left( d(N,j,x) \right)$, with
\begin{align}
d(N&,j,x)  = \frac{x^2}{2} + \ln\Gamma\left(\frac{N}{2}\right) - \ln\Gamma\left(\frac{j}{2}+1\right)  \nonumber \\
& - \ln\Gamma\left(N-j\right) + (N-1-j) \ln\left(\sqrt{2} x \right) - \frac{\ln 2}{2} \nonumber \\
& + \ln\left[1 + (-1)^j \widetilde{\Gamma} \left(\frac{x^2}{2}, \frac{j+1}{2} \right) \right],
\end{align}
and $\widetilde{\Gamma}(\cdot, \cdot)$ denoting the lower incomplete Gamma function
\begin{equation}
\widetilde{\Gamma}(x, a) = \frac{1}{\Gamma(a)} \int_0^x t^{a-1} e^{-t} {\rm d} t\,.
\end{equation}

This way of computing Shannon's \ac{SPB} corresponds to the logarithmic domain approach
proposed in {\cite{Wiechman2008}}, which avoids the numerical over- and under-flows affecting the calculation
of the bound for large block lengths.
%Concerning the logarithmic domain version of the function $f_N(x)$, an alternative, recursive implementation is also presented in \cite{Wiechman2008}.
By considering the parameters of the code used in the $k$-th channel, and by solving $P_{\mathrm{SPB}}(N, \theta, A) = 1- \eta$ with respect to $\gamma$, 
we obtain $\gamma_\eta^{\rm E}(k)$ such that the error probability on that channel is $1-\eta$.

\bibliographystyle{IEEEtran}
\bibliography{biblio}

\end{document}